\documentclass[sigconf,prologue,table,xcdraw]{acmart}

\AtBeginDocument{%
  \providecommand\BibTeX{{%
    \normalfont B\kern-0.5em{\scshape i\kern-0.25em b}\kern-0.8em\TeX}}}

\usepackage{enumitem}
\usepackage[table,xcdraw]{xcolor}
\usepackage{amsmath}
\usepackage{graphicx}
\usepackage{algorithm}
\usepackage{algpseudocode}
\usepackage{soul}
\usepackage{makecell}

\newif\ifnotes
\notestrue

\newcommand{\leilani}[1]{\ifnotes{\leavevmode\color{violet}{\bf (Leilani: #1)}}\else{#1}\fi}
\newcommand{\joanna}[1]{\ifnotes{\leavevmode\color{blue}{\bf (Joanna: #1)}}\else{#1}\fi}

\usepackage{lscape}
\usepackage{subcaption}

\begin{document}

\title{An Adaptive Benchmark for Modeling User Exploration of Large Datasets}

\author{Joanna Purich}
\affiliation{%
  \institution{University of Maryland}}
\email{banana@umd.edu}





\author{Anthony Wise}
\affiliation{%
  \institution{University of Washington}
}
\email{anthw85@cs.washington.edu}

\author{Leilani Battle}
\affiliation{%
  \institution{University of Washington}
}
\email{leibatt@cs.washington.edu}


\begin{abstract}
In this paper, we present a new DBMS performance benchmark that can \emph{simulate} user exploration with any specified dashboard design made of standard visualization and interaction components.
The distinguishing feature of our SImulation-BAsed (or SIMBA) benchmark is its ability to
\emph{model user analysis goals} as a set of SQL queries to be generated through a valid sequence of user interactions, as well as \emph{measure the completion of analysis goals} by testing for equivalence between the user's previous queries and their goal queries. In this way, the SIMBA benchmark can simulate how an analyst opportunistically searches for interesting insights at the beginning of an exploration session  and eventually hones in on specific goals towards the end. To demonstrate the versatility of the SIMBA benchmark, we use it to test the performance of four DBMSs with six different dashboard specifications and compare our results with IDEBench. Our results show how 
goal-driven simulation can reveal gaps in DBMS performance missed by existing benchmarking methods and across a range of data exploration scenarios.

%
%
%
%
%
%
\end{abstract}





\maketitle


\section{Introduction}
\label{sec:introduction}

\begin{figure}[h]
    \centering
    \includegraphics[width=0.5\textwidth]{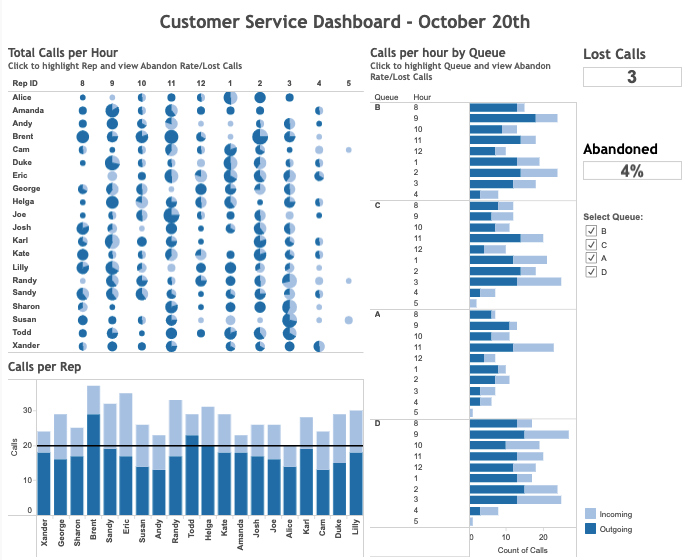}
    \caption{An example dashboard from Tableau Public that enables managers to monitor call centers and respond to issues in employee performance such as abandoned calls.}
    \label{fig:customer_service}
 \end{figure}

Dashboards provide an intuitive collection of visualizations and interaction widgets to help a broad audience of analysts investigate their own questions about a complex dataset~\cite{sarikaya2019}.
Dashboard design is arguably the most pervasive use case for data exploration systems today~\cite{sarikaya2019,bach2022dashboard}, exemplified by the massive investments made by top analytics companies in helping analysts create and maintain interactive dashboards, e.g., by Tableau/Salesforce~\cite{tableau2022tableau,salesforce2022visualize}, Adobe~\cite{adobe2022panels}, and Microsoft~\cite{microsoft2022powerbi}.  To demonstrate the value of dashboards, we walk through a motivating example, illustrated in Figure~\ref{fig:customer_service}.

\begin{example}\label{ex:customer-service-intro}
\textit{The dashboard in \autoref{fig:customer_service} is a real-world example from Tableau Public\footnote{\url{https://public.tableau.com/app/profile/steve.mostello/viz/CustomerServiceDashboard_0/DailyDashboard}}. It tracks the performance of customer service representatives working in a call center.
The total calls per hour and representative are tracked as well as the rate at which calls are abandoned or dropped. A manager might ask a developer to create a dashboard for them to monitor call center performance. The manager might use the resulting} dashboard to identify top performing employees or diagnose an issue that is causing calls to be dropped.
All of the visualizations in the dashboard are linked. For example, if the manager clicks on a bar or pie chart, any visualizations linked to the selected chart must be re-rendered to show the highlighted data, triggering re-execution of the underlying database queries. If the underlying database system is not designed to be interactive, the manager's analysis flow may be disrupted. They may even abandon their exploration out of frustration~\cite{Liu14,crossfilter}.
\end{example}

While developers can quickly create interactive dashboards for end users (e.g., the manager mentioned above) they may struggle to enhance scalability/performance.
Even modest datasets with 10-100 million rows can induce DBMS performance degradation when querying dashboards~\cite{crossfilter}.
%
%
Despite the pervasiveness of dashboards in data exploration and their sensitivity to performance degradation, \emph{\textbf{existing DBMS benchmarks still overlook these interactive use cases~\cite{vogelsgesang2018get}.}} For example, standard analytics benchmarks such as TPC-H~\cite{tpch} and TPC-DS~\cite{tpcds} do not model how end users manipulate visualization interfaces and how these interfaces generate SQL queries in response~\cite{idebench,crossfilter}, hindering their ability to simulate realistic visualization query workloads~\cite{vogelsgesang2018get}.
More dashboard-focused benchmarks support a static set of queries over a fixed set of dashboards (e.g., only bar charts~\cite{crossfilter}, or only a subset of Tableau queries~\cite{ghita2020white}), 
or model end user interactions as purely stochastic~\cite{idebench}.
As a result, dashboard developers and DBMS designers have few options for benchmarking the their systems under more realistic interactive data exploration scenarios.

In this paper, we contribute a SImulation-BAsed (or SIMBA) benchmark for testing the performance of a DBMS with \emph{any dashboard design} composed of standard visualizations and interaction widgets. Given a developer-specified dashboard design in a JSON format, the SIMBA benchmark simulates how the developer's target users might interact with the dashboard to achieve a pre-defined exploration goal. To the best of our knowledge, this paper is the \emph{first} to consider how to generate plausible sequences of user interactions to achieve well-known interactive data exploration goals from the literature. To achieve this, the SIMBA benchmark addresses three challenges in designing benchmarks for interactive data exploration scenarios: (1) expressing user exploration goals such that existing algorithms, e.g., path planning and/or graph search algorithms, can reason about them; (2) translating user interactions into allowable actions that an algorithm can take towards a specified goal; and (3) tracking the algorithm's progress towards achieving a given analysis goal to provide actionable rewards and feedback.

\textbf{Specifying User Goals.} To address the first challenge, developers need a way to specify a user's exploration goals, as well as the allowable actions an algorithm can take to make progress towards these goals. We achieve this through a two-phase process. First, we introduce an algebra for specifying realistic user exploration goals.
Second, we map the algebra to equivalent SQL queries that would provide useful answers to the developer's specified user goal(s).
We demonstrate the utility of our algebra with a collection of reusable goal templates for benchmark users (i.e., developers).
In this way, we can represent user goals as a precise result that can be calculated on the underlying dataset using the corresponding SQL templates.

\textbf{Specifying Allowable Actions to Reach a Goal.} Inspired by previous work~\cite{idebench,2017-vega-lite}, we adopt a JSON-based specification language to specify a dashboard design. We use written specifications to infer an \emph{interaction graph}, i.e., a graph representing what interactions a simulated user can perform and the SQL queries generated by these interactions. Using our interaction graph, we can translate any possible interaction within a given dashboard specification into its corresponding SQL queries over the underlying dataset.

\textbf{Measuring Goal Completion.} Given a navigable space of exploration states, we adapt existing AI algorithms, e.g., path planning and/or graph search algorithms, to identify viable interaction paths towards specified goal states. However, we still need a way to determine whether the algorithm has reached the goal state. To achieve this, we incorporate \emph{a suite of SQL query equivalence functions} for testing whether a generated SQL query matches the specified goal state queries from our templates.

\textbf{Integrating the Three Components to Simulate Goal Directed Exploration.} With these three components, we generate valid sequences of interactions and test whether these interactions will emit the target goal queries. However, analysts rarely march directly towards a pre-specified goal. They opportunistically search their data for interesting paths to pursue and discover meaningful analysis goals along the way \cite{Bat19}. We simulate this oscillation between open-ended to more focused exploration by alternating between Markov models of user exploration behavior and our goal-directed model; we assign probabilities to choosing the actions of either model, and update the probabilities as a simulated user progresses through an exploration session.

\textbf{Evaluation.} To demonstrate the value of our benchmar for developers and DBMS researchers, we test it with four different database management systems (DBMSs) and six dashboard specifications from the literature. We compare the performance of these DBMSs using our benchmark and a competing benchmark IDEBench. Our results show that even when pursuing the same analysis goal, DBMS performance varies across different dashboards due to differences in supported dashboard queries and available data attributes (i.e., data columns). Similarly, differences in exploration goals can also lead to differences in performance outcomes, even when using the same dashboard. Finally, we show that although randomization is useful for simulation-based benchmarking, unconstrained variance can lead to the generation of unrealistic interaction sequences with no clear purpose or analogues in the real world, which the SIMBA benchmark design successfully avoids.

To summarize, this paper makes the following contributions:
\begin{itemize}
    \item We introduce an algebra mapping theoretical goals from the visualization and HCI literature into concrete goal queries that can be tailored to a given dashboard design (\autoref{sec:goals}).
    \item We present an algorithm that leverages SQL queries to simulate how an analyst may interact with a dashboard to achieve a specific goal (\autoref{sec:allowable-actions} \& \autoref{sec:simulation}).
    \item We present the results of running the SIMBA benchmark with six real-world dashboards and four DBMSs, and compare our results with IDEBench. (\autoref{sec:evaluation:exp1}).
    \item We present a user study with six experts comparing logs from SIMBA against real exploration sessions (\autoref{sec:evaluation:user-study}). With appropriate parameters, the SIMBA logs are convincingly realistic and experts see value in the benchmark.
\end{itemize}

All of our data and  code are available at: \url{https://osf.io/vbm8z/?view_only=2e06892f0c104a9e911e8e7599deb2ab}.

\section{Specifying User Goals}
\label{sec:goals}

Query workloads, i.e., observed sequences of DBMS queries, reflect the patterns and cadence of data processing operations made by a target user group. In data exploration, query workloads depend on the sequences of user actions that triggered them, such as clicking on a data point or dragging a slider within a dashboard~\cite{crossfilter,idebench}. Thus, generating meaningful interaction sequences is essential to developing a useful exploration performance benchmark~\cite{idebench}.

However, analysts do not always behave randomly. Like the call center manager, they often have a goal in mind as they explore a dashboard such as answering a question or testing a hypothesis 
\cite{Bat19,lam2018bridging,yan2021tessera, zgraggen2018investigating,alspaugh2019futzing,sarikaya2019}. 
Thus an alternative approach is to generate \emph{goal-directed} simulations.
In this section, we introduce an algebra for formulating data analysis and exploration goals guided by a formative study of the visualization and HCI literature, which we use to generate goal-directed interaction sequences for our simulations.

\subsection{Formative Study of User Exploration Goals}
\label{sec:goals:types}

The visualization community has conducted extensive surveys of user dashboard exploration and design~\cite{bach2022dashboard,sarikaya2019}.
In this section, we describe a formative study which inspired the design of our algebra. Our formative study was conducted in two parts, informed by prior work~\cite{sarikaya2019,Bat19} First, we reviewed five survey papers on dashboard design~\cite{sarikaya2019,bach2022dashboard} and user exploration goals~\cite{Bat19,Brehmer13,lam2018bridging}. These surveys were found by searching the VIS and EuroVis proceedings over the last five years and reviewing the references of identified papers to locate older surveys that were not fully covered by the most recent ones. The second part of our analysis was to search for and analyze relevant dashboards on Tableau Public, informed by the dashboard survey of Sarikaya et al.~\cite{sarikaya2019}. We summarize our findings as follows:

First, we identified four categories of goals relevant to dashboard exploration, summarized by Battle and Heer~\cite{Bat19}
\begin{description}
    \item{\textbf{Understanding Data Correctness and Semantics.} Learning the data schema and checking for data errors, missing fields or otherwise surprising values.}
    \item{\textbf{Characterizing Data Distributions and Relationships.} Understanding the overall shape of the data and identifying potential correlations or outliers for further investigation.}
    \item{\textbf{Analyzing Causal Relationships.} Investigating which attributes (i.e., columns) may drive observed correlations or patterns within the data or may have caused observed outliers.}
    \item{\textbf{Hypothesis Formulation and Verification.} Using observed data relationships to formulate hypotheses about higher-level phenomena and facilitate later decision making.}
\end{description}

Similarly, we noted four common categories of real-world dashboard designs proposed by Sarikaya et al. (decision making, awareness, motivation and learning)~\cite{sarikaya2019}. Guided by their analysis,
we selected a representative set of interactive dashboards for further analysis and generated relevant exploration goals using the categories summarized by Battle and Heer~\cite{Bat19}. We selected two real-world dashboards per dashboard type, found via a search on Tableau Public for active dashboards with publicly available data. Links to all of the dashboards are provided in our supplemental materials~\footnote{\url{https://osf.io/vbm8z/?view_only=2e06892f0c104a9e911e8e7599deb2ab}}.

Our survey review, generation of exploration goals, and interactions with the selected dashboards
yielded two key insights. First, consider the scenario where we achieve goals using only SQL queries (i.e., no dashboard). It appears that \textbf{the majority of user exploration goals from the visualization literature can be achieved using one SQL query or a set of SQL queries}. Furthermore, these queries tend to follow consistent structures, such as frequently grouping by a categorical attribute and aggregating by a quantitative attribute, or pairing two quantitative attributes to assess correlations as in the following example:

\begin{example}\label{ex:example-question}
\textit{To demonstrate our approach, consider our running example in Example~\ref{ex:customer-service-intro} with the customer service dashboard (see \autoref{fig:customer_service}). One of the exploration goals generated for this dashboard was to determine: \textbf{Is there a correlation between call volume and call abandonment?} Using Battle and Heer's categorization, this question is an example of characterizing a data relationship.
We observed that a user could answer this question by clicking/highlighting hours that had a higher call volume and then using the 'Total Calls by Hour' and 'Calls per Rep' visualizations to determine the percent abandoned and if this correlated more closely with the time of day or representative.}
\end{example}

The second insight regards consistencies in dashboard semantics. Although a dashboard may contain a unique combination of data attributes/columns, visualizations, and interactive widgets, \textbf{the underlying SQL queries emitted by the dashboard maintain a consistent structure}. Furthermore, we observe that many interaction types have \emph{overlapping semantics}. For example, checkboxes and radio buttons produce the same underlying categorical filter within SQL queries, likewise for sliders and brush filters. A similar observation is made in recent work on generating dashboards from SQL queries~\cite{chen2022pi2}. We propose a reciprocal idea: \textbf{a dashboard emits certain query structures which constrain the range of exploration goals it can support.}

The insights from our formative study inspired us to create an \textbf{algebra} for capturing user exploration goals and \textbf{mappings} from goal expressions to SQL queries to assess goal attainment.

\begin{table}[]
\caption{Operators for the goal algebra.}
\vspace{-3mm}
\begin{tabular}{c|c|>{\raggedright\arraybackslash}p{0.4\linewidth}}
\hline
\textbf{Operator} & \textbf{Notation}                                              & \textbf{Meaning}                                                                                                                                                                                                                                                                                                                                  \\ \hline
concatenate       & A + B& Place attributes A and B on the same axis.\\ \hline
filter& A - c, 
A - B& Remove all instances of A matching constant c or members of set B.\\ \hline
map& MAP(A,$f_m$), 
$f_m(A)$& To each instance of A apply function $f_m$.\\ \hline
aggregate& AGG(A, $f_a$), 
$f_a(A)$& Calculate aggregate function $f_a$ over attribute A.\\ \hline
compare& B $\times$ A, 
B $\times$ $f_m(A)$& Place attributes A and B on opposing axes. Group by attribute B when comparing aggregates.\\ \hline
\end{tabular}
\vspace{-5mm}
\label{tab:operators}
\end{table}

\subsection{Algebra for Expressing User Goals}
\label{sec:goals:algebra}


Given that the user's exploration goals can generally be achieved via a small set of SQL queries, the goal of the SIMBA benchmark is to manipulate a target dashboard to emit these queries.
In this way, we can easily test whether a dashboard configuration has achieved the desired goal by testing for overlap between the goal queries and the queries emitted by the dashboard during exploration. 
For instance, perhaps the developer was asked to design the dashboard to answer the question:
"What is the average hourly call volume per representative?" The simplest SQL query to answer this question would be of the form \textit{select rep\_id, count(calls) / count(hours) from table group by rep\_id}, but alternative queries could be devised using \textit{AVG()} or by including additional fields in the select statement.

Our formative study seems to suggest that
that only certain types of queries represent valid goals. Therefore, we aim to constrain the space of possible SQL queries for defining suitable goals.
To strike a balance between goal expression and validity, we introduce an algebra for specifying valid exploration goals.

We use the VizQL algebra by Stolte et al. as a starting point~\cite{stolte2002polaris} due to its support for goal-driven data exploration, and adopt the original cross (x), nest (l), and concatenation (+) operators. We introduce dedicated filter (-), map (MAP) and aggregate (AGG) operators to support a wider range of goal definitions (see \autoref{tab:operators}). The filter operator performs element-wise subtraction, removing a non-null elements from a set. The map operator is a user-defined function that maps an element from one domain to a value in the co-domain. Finally, the aggregate operator aggregates all rows with the same value from a categorical field. Below we provide an example for how the algebra might be used to represent user goals.

\begin{table*}[]
\caption{We summarize common expressions in the goal algebra (i.e., well-known user goals from the visualization/HCI literature) as reusable templates, labeled by their relevant data column types (Categorical, Quantitative, and/or Temporal). \vspace{-3mm}}
\scalebox{0.9}{
{\small
\begin{tabular}{l|l|c|ccc}
\textbf{Goal Type}                                                                    & \multicolumn{1}{c|}{\textbf{Generalization}}       &    \multicolumn{1}{c|}{\textbf{Algebra}}                                                                                                                                                         & \multicolumn{1}{c}{\textbf{Cat.}} & \multicolumn{1}{c}{\textbf{Quant.}} & \multicolumn{1}{c}{\textbf{Temporal}} \\ \hline
\rowcolor[HTML]{EFEFEF} 
Analyzing Spread                                                                      & \begin{tabular}[c]{@{}l@{}}Which member of \textbf{{[}categorical attribute{]}} has the largest range/spread of\\ \textbf{{[}quantitative attribute{]}}?\end{tabular}                              & C $\times$ agg(Q)                              & 1                                        & 1                                         &                                       \\
\rowcolor[HTML]{FFFFFF} 
Filtering                                                                             & \begin{tabular}[c]{@{}l@{}}Which \textbf{{[}categorical attributes{]}} have an \textbf{{[}aggregation{]}} of \textbf{{[}quantitative}\\ \textbf{attribute{]}} that is \textbf{{[}comparison operator{]}} \textbf{{[}constant{]}} at any point\\ in time?\end{tabular} & $-$ () & 1+                                       & 1                                         &                                       \\
\rowcolor[HTML]{EFEFEF} 
Finding Correlations                                                                  & \begin{tabular}[c]{@{}l@{}}Is there a strong correlation between \textbf{{[}numerical attribute{]}} and\\ \textbf{{[}numerical attribute{]}}?\end{tabular}                                                             & C $+$ C          &                                          & 2                                         &                                       \\
\rowcolor[HTML]{FFFFFF} 
Identification                                                                        & \begin{tabular}[c]{@{}l@{}}Which \textbf{{[}categorical attribute{]}} consumes the \textbf{{[}max OR min{]}} of \textbf{{[}ordered}\\ \textbf{list of quantitative attributes OR aggregate attributes{]}}\end{tabular}             & C $\times$ (max(Q)$+$min(Q))                  & 1                                        & 1+                                        &                                       \\
\rowcolor[HTML]{EFEFEF} 
\begin{tabular}[c]{@{}l@{}}Measuring Differences\\ Between Group Members\end{tabular} & \begin{tabular}[c]{@{}l@{}}Are there differences in the value of \textbf{{[}quantitative attribute{]}}\\ between the members of \textbf{{[}categorical attribute{]}}?\end{tabular}                                       & C $\times$ agg(Q)        & 1                                        & 1                                         &                                       \\
\rowcolor[HTML]{FFFFFF} 
\begin{tabular}[c]{@{}l@{}}Observing Temporal\\ Patterns\end{tabular}                 & \begin{tabular}[c]{@{}l@{}}How does change in \textbf{{[}temporal attribute{]}} affect patterns in \textbf{{[}quantitative}\\ \textbf{attribute OR aggregate attribute{]}}, if at all?\end{tabular}                        &  DAY(T) $\times$ agg(Q)               &                                          & 1                                         & 1                                    
\end{tabular}}
}
\label{tab:templates}
\end{table*}

\begin{example}
\emph{Continuing our running example, the call center manager might have the goal of answering the following question using the developer's dashboard: is there a correlation between call volume and call abandonment?
To analyze this, we are interested in calculating call volume (total calls over time), and call abandonment (sum of call abandonment events over time). We need a temporal variable (T) to measure time and we want to compare call volume (C) versus call abandonment (A). Using our algebra, this would be represented as: T x count(C) x sum(A).}

\emph{These variables map to rep\_id (R) and call volume (C) as the total call volume and the total number of calls for each representative, respectively. Expressing this using our algebra goals (see Table 1), we get $AGG(C, f_{sum})$ and $AGG(C, f_{count})$. We  then can go further by using $MAP(\frac{AGG(C, f_{sum})}{AGG(C, f_{count})}, f_{avg})$, then finally for each representative $R \times MAP(\frac{AGG(C, f_{sum})}{AGG(C, f_{count})}, f_{avg})$.}
\end{example}

\subsection{SQL Translation and Example Goals}
\label{sec:goals:templates}

The goal of the SIMBA benchmark is to allow benchmark users to specify their own exploration goals through the algebra, which in turn facilitates translation into appropriate SQL queries.
%
%
We illustrate how the algebra maps to SQL with an example:

\begin{example}\label{ex:example-query}
\textit{Consider the analysis question presented in Example~\ref{ex:example-question}. This question asks about a relationship between two numerical variables: total calls (\texttt{COUNT(*)}) and total abandoned calls (\texttt{SUM(abandoned)}). Thus, this question can be generalized as:} ``Is there a strong correlation between \textbf{numerical attribute 1} and \textbf{numerical attribute 2}?'' \textit{Given the question is parameterized by data type, and assuming we have access to the underlying dataset schemas for the dashboards, this question can be applied to any dashboard that contains at least two numerical fields (or in this case, aggregates that produce numerical values). Going one step further, this question template can be captured through the following SQL template:}
\begin{verbatim}
SELECT [numerical attribute 1], [numerical attribute 2]
FROM [table]
\end{verbatim}
\textit{To account for the volume of calls over time, we can include a modulating variable, for example calculating calls per \texttt{hour} rather than simply total calls, as a grouping variable:}
\begin{verbatim}
SELECT [modulator], [numerical attribute 1],
    [numerical attribute 2]
FROM [table] GROUP BY [modulator]
\end{verbatim}
\textit{When we populate the template for our call volume example, we get the following SQL query representing our target goal query:}
\begin{verbatim}
SELECT hour, COUNT(*) as call_volume,
    sum(abandoned) as call_abandonment
FROM customer_service GROUP BY hour
\end{verbatim}
\end{example}


However, it may be difficult for first-time users to formulate realistic user exploration goals without examples. 
Further, there are no guarantees that arbitrary SQL queries will reflect valid exploration goals as defined in the visualization literature. Our algebra is designed to address this problem.
To further ease the process of using our benchmark,
we present six reusable algebra expression templates that cover a diverse set of exploration goals, shown in \autoref{tab:templates}. Our expressions cover both the exploration goals categorized by Battle and Heer~\cite{Bat19} 
and dashboard types categorized by Sarikaya et al.~\cite{sarikaya2019}
(see \autoref{sec:goals:types}).
For example, the question in Example~\ref{ex:example-question} corresponds to row four of \autoref{tab:templates} (Finding Correlations).
Further, each expression is pre-translated into corresponding SQL queries, facilitating their comprehension by benchmark users.

Although our goal expression templates are designed to cover most benchmark users' needs, they are not exhaustive. For example, similar to prior work in this area~\cite{idebench,crossfilter}, most of our selected dashboards apply binned aggregation to temporal rather than quantitative attributes. However, these expressions can easily be extended to support alternative scenarios by swapping out or adding the desired data attributes, e.g., swapping temporal for quantitative or categorical attributes to calculate bins. Further, users can always form their own goals using the algebra instead. Note that using these templates and the algebra are optional; any SQL query supported by SIMBA's simulation code can be used to specify user goals instead of algebra expressions if desired.

\section{Specifying Allowable Actions}
\label{sec:allowable-actions}

Given a specified end user goal, the next step is to detect which dashboard interactions would bring a simulated end user towards this goal. 
To enable SIMBA to infer what interactions are allowed within a dashboard and how they translate to SQL queries, we provide a language for developers to \emph{specify} \emph{any dashboard design} consisting of common visualization and interaction components.
Our specification language merges existing specification formats from IDEBench~\cite{idebench}, Polaris/Tableau~\cite{stolte2002polaris}, Vega-Lite~\cite{2017-vega-lite}

\begin{figure*}[t]
    \centering
    \includegraphics[width=0.8\textwidth]{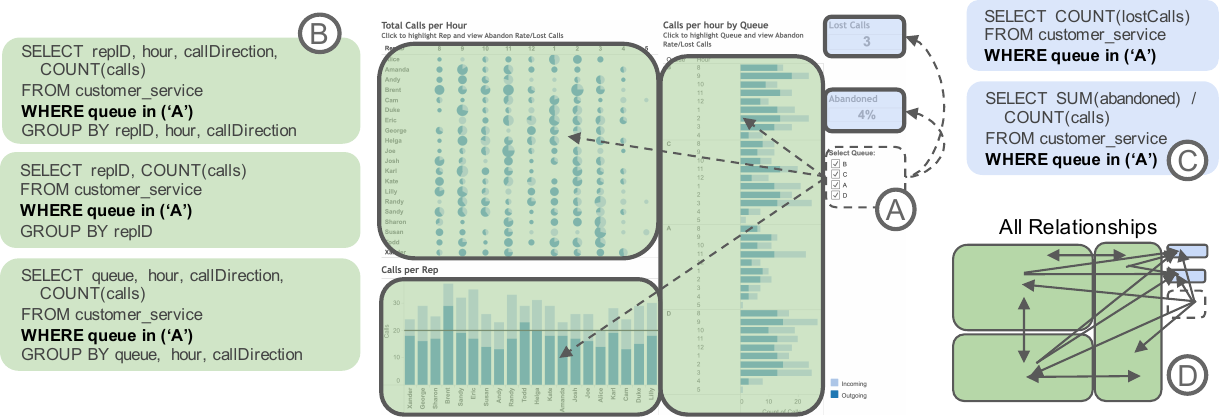}
    \vspace{-2mm}
    \caption{(A) Shows how relationships between dashboard visualizations and interaction widgets are captured in the SIMBA graph layer. Here, the checkbox widget of the Customer Service Dashboard can trigger updates in the main visualizations in green and summary statistics in blue. These updates trigger new SQL queries in the data layer to be executed by the DBMS, shown in (B) and (C). Finally, (D) shows all of the relationships in the graph layer for the customer service dashboard.}
    \label{fig:layers_update}
    \vspace{-3mm}
\end{figure*}

Similar to prior work~\cite{idebench,2017-vega-lite}, our specification language relies on three components (dataset, visualization, and interaction specifications)
to simulate end user interactions.
These inputs are used to form a \emph{graph data structure}, which enables us to reason about any possible interaction an end user might perform on this dashboard by maintaining a \emph{joint representation} of the dashboard state, i.e., the state of the system in terms of end user interactions performed in the dashboard interface (\textbf{Interaction Layer}) \emph{and} the state of the system in terms of the underlying database schema and SQL queries issued to the DBMS (\textbf{Data Layer}).
For sake of space, we briefly summarize this language at a high level, explain how we translate dashboard specifications into our graph representation, and discuss how we map user interactions (i.e., manipulations on the graph) to their corresponding SQL queries. Please see our supplemental materials for detailed example usage.

\subsubsection{Specification Language}
\label{sec:allowable-actions:spec-language}





This language has three components: the Database, Interface, and (optional) Interaction Specification. 
The Database Specification is inherited from IDEBench~\cite{idebench} and expresses the
dataset in a way that can be easily ported across various DBMSs.
%
The Interface Specification extends both IDEBench~\cite{idebench} and Vega-Lite~\cite{2017-vega-lite} to make it easier to specify complete dashboards (multiple visualizations and interaction widgets) rather than just a single visualization.
It defines how visualizations and interaction widgets interconnect, such as how adjusting a slider interaction widget might refine the data displayed in a bar chart visualization
(see \autoref{fig:layers_update} for examples).

\subsubsection{Generating the Interaction Layer}
\label{sec:allowable-actions:graph-layer}

The generate the interaction layer (i.e., the user interface side of our joint representation),  we initialize each visualization and interaction widget in the specification. Then, we draw a directed edge from each source visualization/interaction widget to a target visualization/interaction widget if interaction with the source precipitates a change in the target. For example, in the case of a filtering dropdown menu interaction widget, an edge would be drawn from the dropdown menu to the visualization that it filters. 
The interaction layer supports two types of interactions: \emph{data manipulations}, 
which execute an existing interaction widget in the dashboard (i.e., just uses the dashboard user interface as-is)
\emph{interface manipulations} which modify the original dashboard definition (i.e., alters the dashboard's user interface, for example, to add/remove a visualization).

\subsubsection{From Interactions to SQL Queries}
\label{sec:allowable-actions:interactions-to-sql}

For each simulated data manipulation interaction, the SIMBA benchmark propagates SQL filter statements across the dashboard's graph layer for each node that can be reached via directed edges. The additional filter statements added to each visualization alter the data displayed. Therefore, we must update the associated SQL query to fetch these results from the DBMS using the data layer. The data layer represents each node as a SQL query where the node properties in combination with the Database Specification define how the SQL query is generated.

In the graph representation, as nodes are initialized for each visualization in the interaction layer, these nodes are also passed to the data layer.
The data columns specified in each visualization are
validated against the Database Specification to identify their parent table or custom field definition. To build the corresponding SQL query, the validated columns are added to the SELECT statement and their parent tables are joined according to the Database Specification and added to the FROM statement. For example, the SQL query for the \emph{Lost Calls visualization would be:}

\begin{verbatim}
SELECT COUNT(lostCalls)
FROM customerService;
\end{verbatim}


When a data manipulation interaction is evaluated (i.e., the end user manipulates an interaction widget like a slider), we analyze the affected nodes in the graph representation to obtain the set of SQL queries that would be executed when this interaction occurs.

\begin{example}
\emph{Let us revisit our customer service example. All possible interactions and their impact on various visualizations are captured in the graph layer, shown in \autoref{fig:layers_update}D. Suppose we perform the \emph{data manipulation interaction} indicated in \autoref{fig:layers_update}A, where selecting
the ``\texttt{queue A}'' checkbox interaction widget filters all five visualizations in the dashboard.
To map this filter statement to SQL queries, we first locate the source of the interaction, the checkbox node, and update the node properties to reflect that it has been filtered down to ``\texttt{queue A}.'' Next, we recursively propagate this filter across all outbound edges of the checkbox node. For each target node connected by an edge, we add the filter condition to the list of node properties.}

\emph{At the conclusion of the ``\texttt{queue A}'' checkbox interaction there are five visualization nodes with property updates, as indicated by dashed lines in \autoref{fig:layers_update}A. For each of these nodes, an updated SQL query needs to be generated in the data layer using the corresponding SQL filter statement. 
Therefore, we update the \textit{WHERE} clause in each query to include the statement ``\texttt{AND queue = 'A'}'' in the data layer. These updates are triggered automatically by the  SIMBA benchmark.}
\end{example}

\section{Generating Exploration Sessions}
\label{sec:simulation}

With the ability to specify end user exploration goals and valid dashboard interactions in pursuit of these goals, it is possible to generate sequences of interactions to achieve a specified goal. In this section, we explain how sequences of goal-directed data exploration interactions are generated. However, analysts rarely know exactly how to reach a goal directly, and may even shift their goals over time~\cite{Bat19}. Thus, we also explain how we augment our goal-directed interaction sequences to reflect more realistic exploration scenarios.

\begin{figure}
    \centering
    \includegraphics[width=0.9\columnwidth]{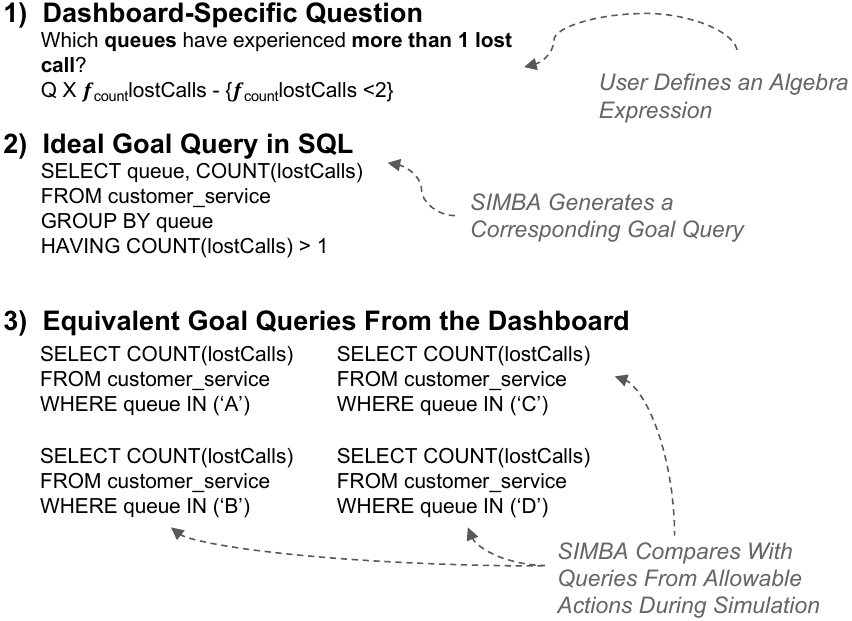}
    \caption{The Analyzing Spread template populated for the Customer Service dashboard. In this case the goal query is not syntactically achievable, but it is semantically achievable as the union of four queries generated by the Lost Calls visualization and Queue Checkbox interaction widget. Therefore, the parameterized goal can be achieved if all four queries are generated, in any order, during an exploration session.\vspace{-5mm}
    }
    \label{fig:analyzing_spread}
 \end{figure}
\subsection{Simulating Targeted Exploration}
\label{sec:simulation:targeted}

Targeted exploration happens when the user has a well-formed analysis goal in mind~\cite{Bat19,zeng2022evaluation}. To execute a targeted simulation, the SIMBA benchmark requires a dashboard specification
(as stipulated in \autoref{sec:allowable-actions:spec-language}) and a sequence of goal queries to be achieved
(as stipulated in \autoref{sec:goals:templates}). We refer to a set of goal queries as a \emph{goal set}.
We generate a targeted exploration session by navigating the interaction layer to 
emit queries that can be compared against the goal set.
With each step of the simulation, we check our work in the data layer of the graph representation to test if any queries in the goal set have been fulfilled.

As suggested by our formative study in \autoref{sec:goals}, users do not blindly explore, they iteratively set analysis goals, explore until they find an answer, and then set new goals. We represent this goal-driven behavior as a sequence of interactions to achieve each goal query within the goal set.
In the targeted exploration setting we design an end user model that we refer to as the Oracle, which takes a shortest path approach to achieving the goal set. 
%
%
%
Specifically, the Oracle model is given the goal set and the interaction layer of the graph representation (i.e., all possible interaction paths and corresponding dashboard states) as input and solves for the shortest sequence of interactions that will achieve the goal set.
When multiple goals are viable next steps, the Oracle model prioritizes interactions
that achieve the goal queries in fewer interactions. \textbf{We formulate the Oracle as a state-space planning problem.}

\subsubsection{Generating Immediate Next Steps}
Given the current state of the interaction layer, the Oracle model must decide the next interaction to accomplish the current goal set. The Oracle model leverages the graph representation by searching
the interaction layer and checking the SQL queries in the data layer to determine if analysis goals have been met. For example, in \autoref{fig:goal_to_queries} an Analyzing Spread goal is executed against the Customer Service dashboard. To generate the next interaction, we search the interaction layer to identify all possible interactions. In this case, there are four viable interaction widgets.
For example, there are 16 allowable configurations of the checkbox interaction widget, and each of the visualizations contain embedded interaction widgets for manipulating various combinations of data columns, e.g., representative, hour, queues, or incoming vs. outgoing calls. The optimal next interaction should bring us closer to achieving the goal set, e.g., closer to tallying the lost calls observed for each call queue.

\subsubsection{Measuring Goal Completion}
\label{sec:simulation:goal-completion}

We consider a target goal query solved if the result set of this query is ``covered'' by the result sets of all simulated SQL queries thus far. In other words, if at some point in the simulation the simulated end user saw the desired records specified by the goal query, then this goal has been achieved.

\begin{figure*}[h]
    \centering
    \includegraphics[width=\textwidth]{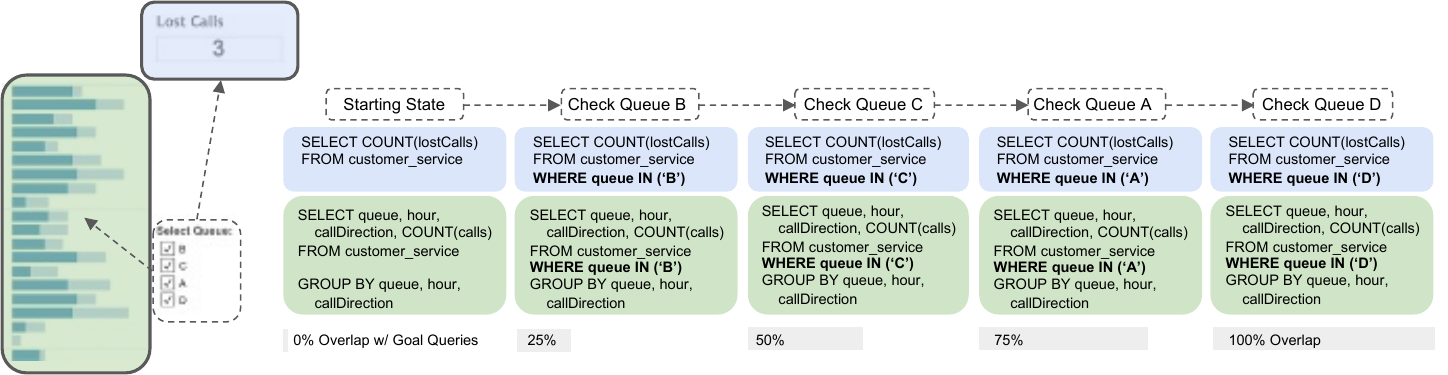}
    \vspace{-6mm}
    \caption{An example workflow, i.e., interaction sequence, on a sub-graph from the Customer Service dashboard and the Analyzing Spread template from \autoref{fig:analyzing_spread}. As the user selects various queues, the properties of the connected nodes are updated and new queries are generated. In this case, the goal is achieved, i.e., the goal queries are covered, in four interactions.
    }
    \label{fig:goal_to_queries}
    \vspace{-1mm}
 \end{figure*}

This is a comparison of two results sets. The first set represents the anticipated query results from the specified goal query, which we denote as $R_g$. If there are two goals to achieve $g$ and $g'$, we can represent this as a union of the result sets across the corresponding goal queries: $R_g \cup R_{g'}$, which also extends easily beyond two goals to generalize for goal sets. The second results set is the set of observed query results for each simulated interaction thus far. For example, if only one interaction has been simulated $i$ (e.g., selecting queue \texttt{B in \autoref{fig:goal_to_queries})}, then there is only one result set to compare, denoted as $R_i$. If two interactions have been simulated $i$ and $i'$ (e.g., selecting queue \texttt{B, then \texttt{C})}, we denote this as $R_i \cup R_{i'}$. Given the goal set $G$ and the set of all simulated interactions $I$, we consider all target goal queries solved if:
\begin{equation*}
    \bigcup_{g \in G} R_g \subseteq \bigcup_{i \in I} R_i
\end{equation*}

We determine the equivalence of query result sets in three ways: (1) \emph{syntactical equivalence}, or seeing if the text of the queries is the same, (2) \emph{semantic equivalence}, or reasoning about what the queries represent to see if the results should be the same, or (3) \emph{result equivalence}, or executing the queries and inspecting the results to test if they overlap. We can use similar techniques to test for query subsumption, for example if one query is the prefix of another query in terms of text, or if semantically one query should subsume the other. We describe our equivalence/subsumption methods below.

 \paragraph{Syntactical Equivalence.} A query is syntactically equivalent to the goal query if the query's text covers at least the same columns and rows as the goal query's text. 

\paragraph{Semantic Equivalence.} We use the SPES SQL solver~\cite{zhou2020spes} to infer semantic equivalence between goal queries and dashboard interaction queries. 
Two queries are equivalent if they produce the same results given any valid input relations, which SPES infers by compiling SQL queries into denotational semantics and using mathematical principles to reason about them.

However, there are some SQL features that SPES does not support such as expressions in \verb|GROUP BY| clauses, aggregate functions such as \verb|MAX()| and \verb|MIN()|, and functions that extract values from existing data such as \verb|YEAR()| and \verb|MONTH()| from a date. For this reason, we extend SPES with two new methods for equivalence testing: result set overlap and string matching. For example, we infer equivalence if SPES returns a match or if string matching indicates >95\% similarity after processing to remove additional whitespace.
 
 \paragraph{Result Equivalence.} Query results are considered equal when the goal query result is subsumed by the query results represented in the current dashboard state. Specifically, every column and row in the goal result set must also be present in the dashboard result set, but the dashboard may contain more rows or columns than what is stipulated in the goal query. For example, if the goal result is a subset of the query results rendered in a specific visualization component, then result equivalence can be detected in the dashboard.

 \paragraph{Measuring Progress.} Given the ability to determine SQL equivalence or subsumption, goal completion becomes straightforward to detect. However, making progress towards a goal is not the same as completing it. We measure progress toward a specified goal as \emph{overlap} between result sets rather than subsumption or equivalence. The more results are ``covered'' by a potential interaction (i.e., the more results overlap with the anticipated result set), the closer it brings the simulation to reaching the target goal state. In this way, the Oracle model can compare the utility of potential next steps by comparing overlaps produced by their corresponding query result sets (i.e., maximizing $\bigcup_{g \in G} R_g~~~\bigcap~~~\bigcup_{i \in I} R_i$). 
 
The overlaps define the heuristic function for the LookAhead forward planner~\cite{ghallab2016automated} to determine the (inter)action plan to achieve the goal (see Algorithm~\autoref{alg:goal-simulation}). $S$ is the set of all possible dashboard states in the graph representation. $s_n$ represents the state achieved after the $n$th interaction has been applied. LookAhead is initialized with the current state $s_0$. We then identify all graph states $s_{n+1}$ that satisfy $Applicable(s)$, i.e., the states that can be reached by applying interaction $a$ to state $s_0$. If $s_{n+1} \in S$ then $a$ is a valid interaction.

While there exist $Applicable$ states, we recursively invoke Lookahead to maximize the heuristic function $\theta$ to find the best set of future interactions. $\theta$ calculates the overlap of the query result sets for each state and we choose the interaction that maximizes $\theta$ to add to the policy $\pi$. We then execute the first step in the plan, record the new state $s_{n+1}$ of the graph representation and recursively call Lookahead to re-plan. This iterative process continues until the goal state is achieved or no further actions are possible.

Note that dashboard developers can specify user goals directly in SQL rather than the algebra in \autoref{sec:goals:algebra, as long as the queries are supported by SIMBA's SQL equivalence tests.}

\begin{algorithm}
\caption{Lookahead-Partial-Plan$ (\Sigma, s_0, S)$}\label{alg:goal-simulation}
\begin{algorithmic}[1]
\State $s \gets s_0$
\While{$Applicable(s) \neq \emptyset$}
\State $\pi \gets Lookahead(s,\theta) $ \Comment{Planning}
\If{$\pi = \emptyset$ then return failure}
\Else{}
\State perform partial plan $\pi$ \Comment{Acting}
\State $s \gets$ observe current state
\EndIf
\EndWhile
\Procedure{Applicable}{$s_0$}
\State $s_1 \gets$ Apply action $a_1$ to state $s_0$
\State \textbf{return} $s_1 \in S$
\EndProcedure
\Procedure{$\theta$}{$s, R_g$}
\Comment{Result set achieved by state $s$}
\State $R(s) \gets$ query result set achieved in state $s$
\State \textbf{return} $\bigm|R_g \cap R(s)\bigm|$
\EndProcedure

\end{algorithmic}

\end{algorithm}
\vspace{-5mm}

\subsection{Simulating Open-Ended Exploration}
\label{sec:simulation:random}

Unlike targeted exploration,
open-ended exploration is often characterized by a lack of clear goals and spontaneous interactions (see \autoref{sec:goals}).
We adopt the definition of open-ended exploration used by Battle and Heer~\cite{Bat19} and Zeng et al.~\cite{zeng2022evaluation}.
Contrary to the Oracle model, which takes the shortest path to achieve the goal set, open-ended exploration allows end users to perform unexpected interactions such as to learn new information or make mistakes.

We extend the stochastic approach of Eichmann et al. to simulating open-ended exploration~\cite{idebench}.
by initializing the Markov chain, randomly selecting an initial interaction and then applying it to the dashboard graph.
Specifically, given the current state of the dashboard, we randomly select an outgoing edge from the corresponding node of the graph representation. The probability of selecting an edge is determined from how often analysts selected this type of interaction in prior user studies~\cite{idebench.}
We repeat this random selection process until the goal set has been achieved
as determined by testing for query equivalence (see \autoref{sec:simulation:goal-completion}).


\paragraph{Predicting Interaction Parameters.} Once an interaction has been predicted, the next step is to apply the effects of this interaction to the dashboard state, which may require additional parameters. For example, to perform a checkbox selection interaction 
in the Customer Service dashboard,
the simulator must decide which of the
four call center queues (i.e., four chechboxes),
will be checked or unchecked.
Once the selection interaction has been predicted, the simulator fills in these parameters using uniform probabilities.
This approach extends easily to other interaction types, such as randomly selecting the start and end points for dragging a slider.
Note that we simulate serial manipulation of interaction parameters because users can only perform one click at a time. For example, users cannot physically click on two checkboxes simultaneously.


To simulate open-ended exploration in a range of scenarios, we provide a library of pre-set transition probabilities, including probability distributions originally proposed by Eichmann et al. for IDEBench~\cite{idebench}. Users can also specify their own custom transition probabilities as input to the SIMBA benchmark.

\subsection{Interweaving Targeted and Open-Ended Exploration}
\label{sec:simulation:combined}

During exploratory visual analysis, users often alternate between targeted exploration
and open-ended exploration~\cite{wongsuphasawat2017voyager,Bat19}.
In the beginning, the user may not have any concrete goals~\cite{alspaugh2019futzing,Liu14}, making unstructured interactions more likely. However, as the user learns more about their dataset, they become more focused on answering specific questions~\cite{zgraggen2018investigating}. In this section, we describe how we alternate between targeted and open-ended interaction sequences over time.

\paragraph{Modeling Shifts in Exploration Strategy.} We use the Oracle model to simulate targeted exploration to achieve developer-specified goal queries (see \autoref{sec:simulation:targeted} for details) and a Markov model approach extending the work of Eichmann et al.~\cite{idebench} to simulate open-ended exploration (see \autoref{sec:simulation:random} for details). We alternate between targeted and open-ended exploration by switching from one model to another to simulate the end user's next interactions. In this way, we can control when and for how many interactions a certain exploration strategy is adopted by the simulation, which we achieve by assigning probability distributions to the models. For example, a distribution of 0.5 means it is equally likely for the simulation to choose between the two models. A distribution of 0 or 1 means that only one model will ever be selected.

\begin{figure}[t]
    \centering
    \includegraphics[width=0.6\columnwidth]{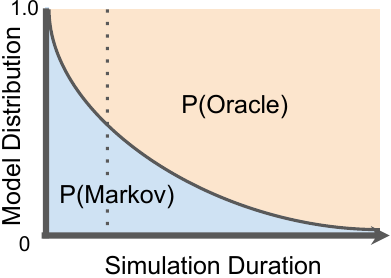}
    \vspace{-3mm}
    \caption{We use exponential decay to shift simulations from being exploration- (Markov model) to goal-focused (Oracle model). At the dotted line, both models are equally likely.\vspace{-5mm}}
    \label{fig:simulation-distribution}
\end{figure}

Each exploration session is initiated as an open-ended exploration session dominated by the Markov model.
However, as the simulation progresses, the ``end user'' learns more about their data and becomes more focused, decreasing the probability of selecting the Markov model. We implemented this behavior as a decision function that applies exponential decay to the likelihood of selecting the Markov model, depicted in \autoref{fig:simulation-distribution}.
We tune our decision function with default parameters that yield interaction session durations consistent with those observed in prior studies~\cite{idebench,crossfilter,Bat19,zgraggen2016progressive,zgraggen2018investigating}.
These parameters may also be used to simulate the user's familiarity with the dashboard or data. For example, experienced users know what they are looking for and thus may quickly default to following the shortest path (i.e., a lower initial probability assigned to the Markov model and faster decay).
However, novice users are better modeled with a lower decay rate, resulting in spending more time in the open-exploration phase.

\paragraph{Modeling Shifts in Exploration Goals.} Analysts often target more than one goal as they explore a dataset~\cite{Bat19}. We can simulate this behavior by  imposing an order on the goal set as input to the benchmark.
In this case, the simulation will run using the first goal query and the default dashboard state as its initial input. When the first goal query is achieved, the simulation will repeat the same process but use the current dashboard state and the second goal query as input. Then, this process is repeated until all goal queries have been achieved.
In this way, we can model a progression of exploration through goal query transitions.
To streamline the process, we have provided three default goal orderings, each representing well known goal orderings from the literature~\cite{Bat19, shneiderman96eyes, crossfilter}. These defaults can easily be copied and augmented to generate similar orderings for different input datasets and dashboard specifications (please see our supplemental materials for examples). In the future, we plan to extend the benchmark to dynamically generate goal orderings based on the current model and dashboard states.



\section{Comparing With Other Benchmarks}
\label{sec:comparison}




\paragraph{Comparing with TPC-H and TPC-DS}
TPC-H and TPC-DS are ``decision support'' benchmarks from the Transaction Processing Performance Council (TPC)~\cite{tpch,tpcds}. They reflect common business operations for DBMSs. However, as noted by prior benchmark creators~\cite{crossfilter,idebench}, these TPC benchmarks do not reflect real-time end user interactions with dashboards. They are closer to routine reports generation and updates to existing monitoring systems. Thus, they are not considered accurate reflections of interactive data exploration scenarios by the visualization and HCI research communities~\cite{battle2017benchmark,battle2018evaluating} (or even the database community, in some cases~\cite{vogelsgesang2018get}).

\paragraph{Comparing with the Crossfilter Benchmark} The Crossfilter Benchmark measures the performance of DBMSs under rapid, real-time query generation by a specific type of dashboard~\cite{crossfilter} (crossfilter dashboards~\cite{moritz2019falcon}). While the Crossfilter benchmark aims to achieve similar goals as the SIMBA benchmark, it is a static benchmark. While the benchmark creators did collect real user data, DBMSs can only be tested using the three crossfilter dashboards provided in the benchmark, and only with the limited user data collected by the benchmark creators from their user study. In comparison, the SIMBA benchmark allows developers to test any dashboard design made of standard visualizations and interaction widgets. Further, SIMBA can simulate an unbounded number of exploration sessions with any specified dashboard, allowing for more variation and stress testing in terms of total users simulated and dashboards tested. Note that the SIMBA, Crossfilter, and IDEBench benchmarks all measure the same evaluation metrics and provide support for approximate visualization, so they are equivalent in terms of metrics.

\paragraph{Comparing with IDEBench}
Similar to the SIMBA benchmark, IDEBench was developed to test the performance of DBMSs by simulating end user interactions with a developer-specified dashboard~\cite{idebench}. In fact, our open-ended exploration model is an extension of the original Markov chain design proposed by Eichmann et al. for their simulations~\cite{idebench} (see \autoref{sec:simulation:random}). However, IDEBench simulates user exploration as a fully stochastic process. In other words, end users are simulated as behaving randomly, where the randomness is controlled somewhat by varying the probability of different types of interaction widgets.

However, analysts do not always behave randomly, as suggested by our formative study (see \autoref{sec:goals)}. Analysts often have a goal in mind as they explore a datset~\cite{Bat19,lam2018bridging}. Unlike the SIMBA benchmark, IDEBench does not consider end user goals, limiting its ability to produce realistic simulations of end user analysis behavior. Furthermore, the realism of DBMS benchmark simulations has never been measured. In contrast, the SIMBA benchmark contributes: an algebra for concise expression of end user exploration goals, a model for simulating goal-directed exploration of a developer-specified dashboard, and a method for interleaving open-ended and goal-directed (i.e., targeted) models of end user exploration behavior to generate more realistic simulations. In the next section, we evaluate the SIMBA benchmark through a synthetic comparative analysis with IDEBench and a user study with real-world analysts and database/visualization experts. Our results show that a purely stochastic simulation method has some drawbacks for benchmarking DBMS performance in visualization and exploration contexts.

\section{Evaluation}
\label{sec:evaluation}

\begin{figure*}[t]
     \centering
     \begin{subfigure}[t]{0.3\textwidth}
         \centering
         \includegraphics[width=0.9\textwidth]{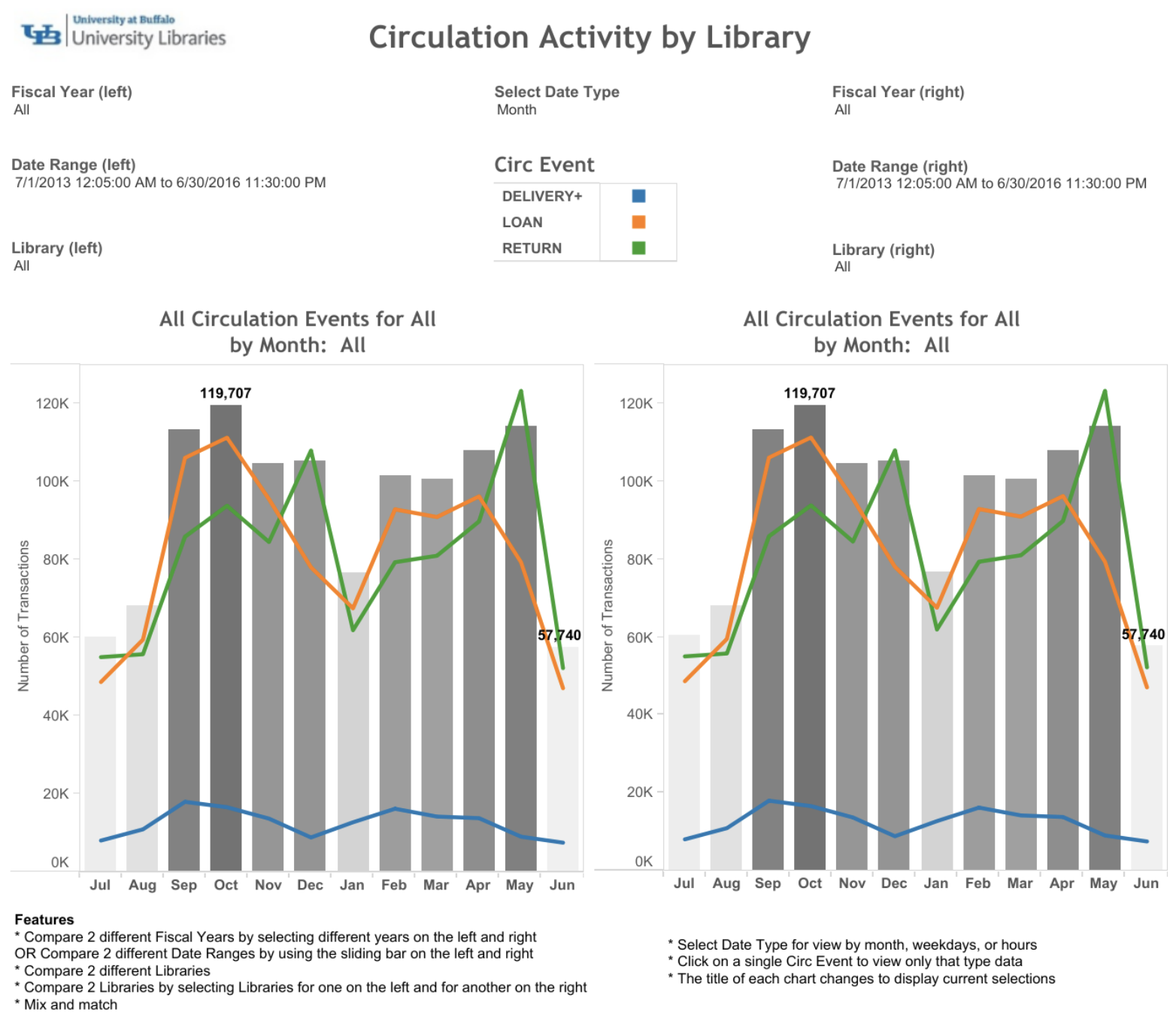}
          \caption{Circulation Activity by Library (2Q, 2C)}
         \label{fig:circactivity}
     \end{subfigure}
     \hfill
     \begin{subfigure}[t]{0.3\textwidth}
         \centering
         \includegraphics[width=\textwidth]{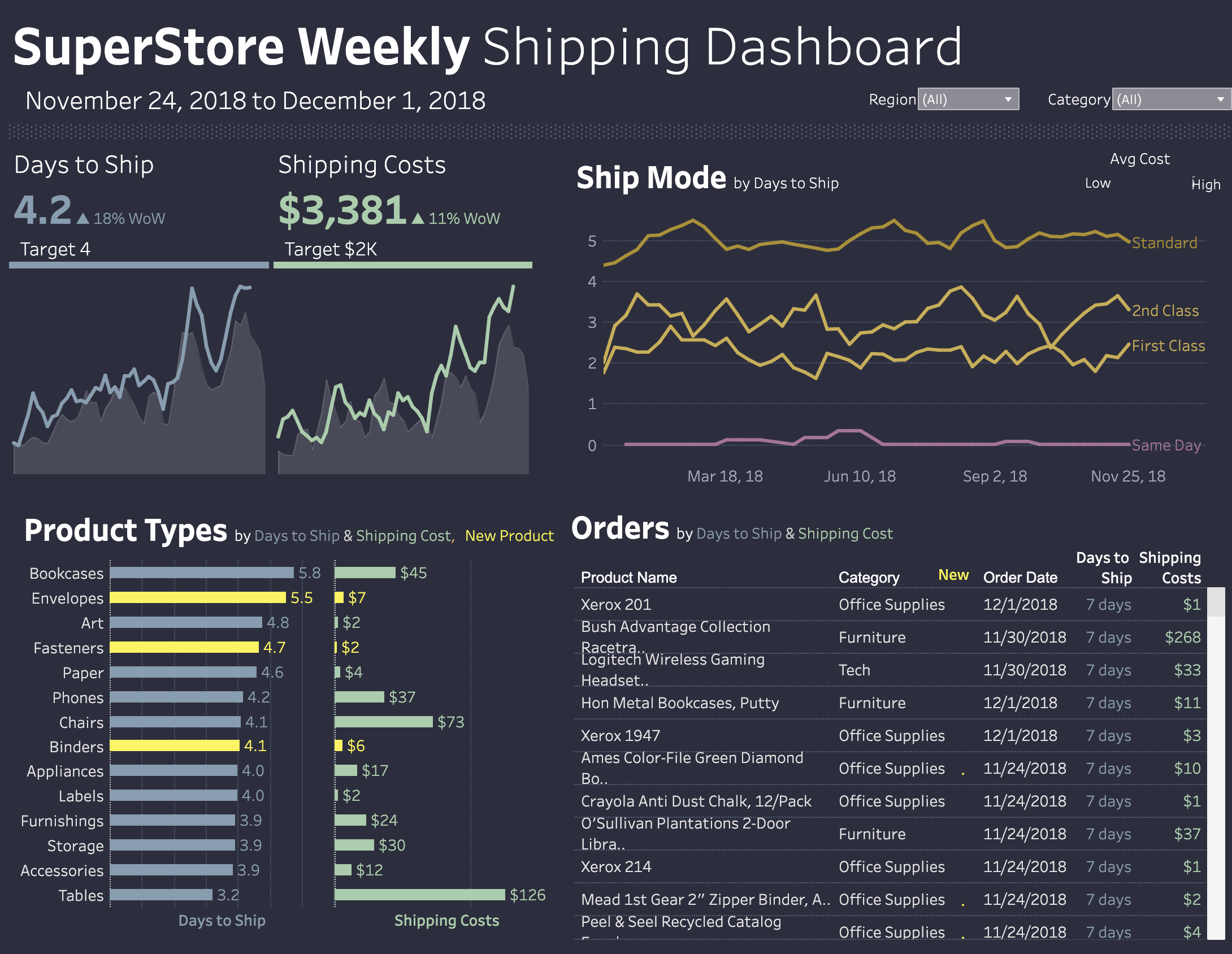}
          \caption{Supply Chain (5Q, 18C)}
         \label{fig:superstore}
     \end{subfigure}
     \hfill
     \begin{subfigure}[t]{0.3\textwidth}
         \centering
         \includegraphics[width=\textwidth]{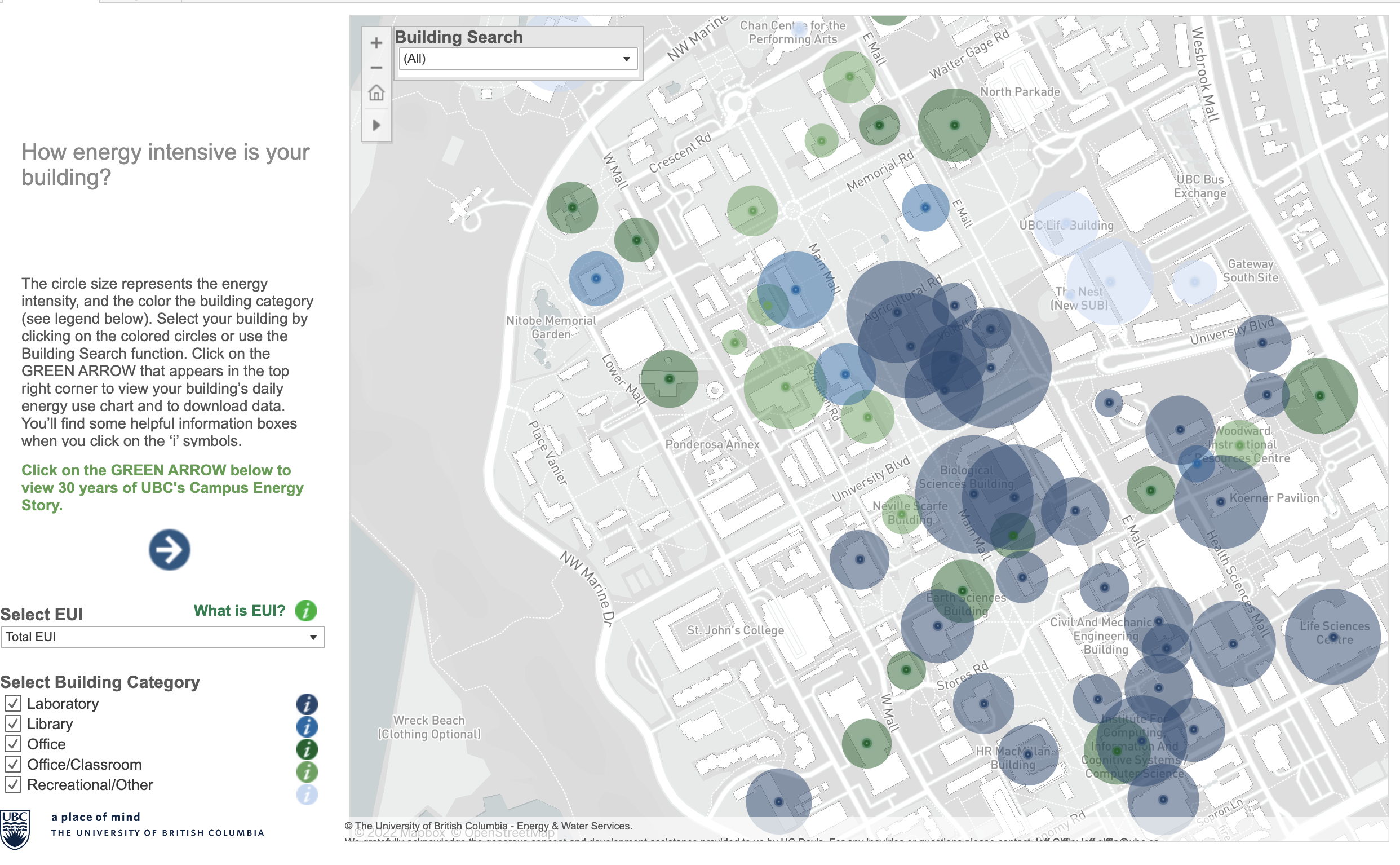}
          \caption{UBC Energy Map (22Q, 4C)}
         \label{fig:campusenergymap}
     \end{subfigure}
     \hfill
     \begin{subfigure}[b]{0.3\textwidth}
         \centering
         \includegraphics[width=\textwidth]{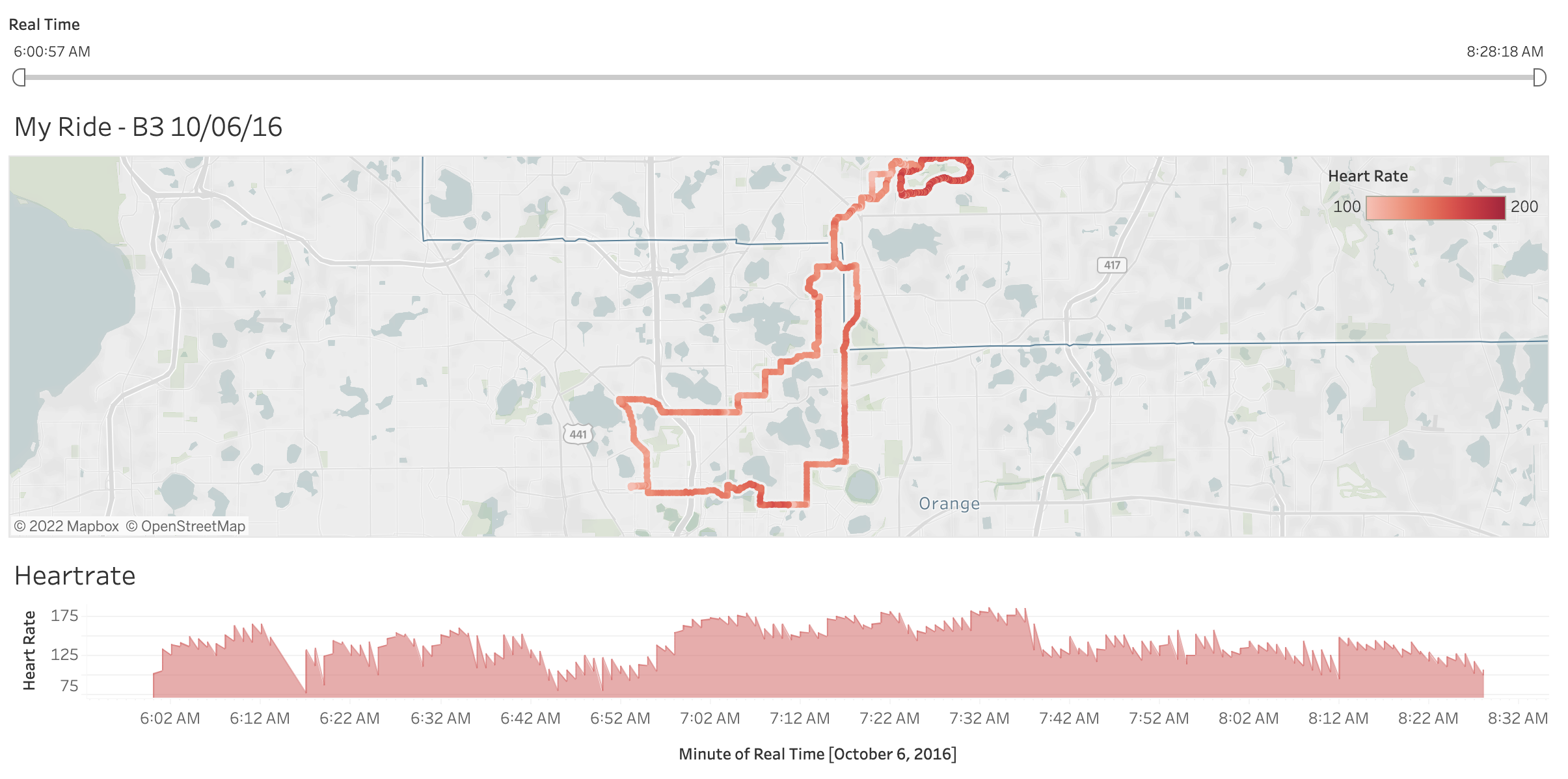}
          \caption{MyRide (10Q, 3C)}
         \label{fig:myride}
     \end{subfigure}
     \hfill
     \begin{subfigure}[b]{0.28\textwidth}
         \centering
         \includegraphics[width=\textwidth]{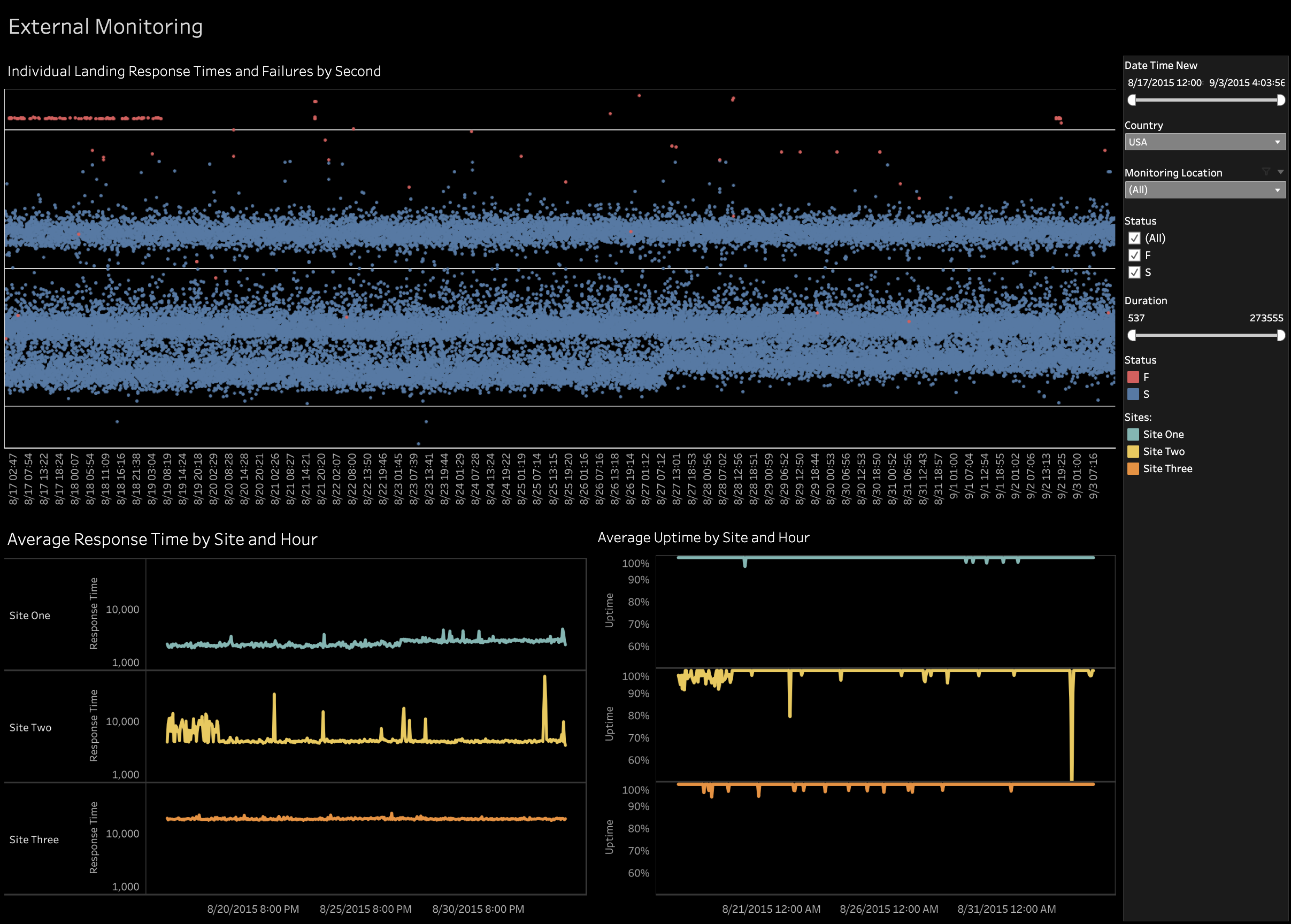}
          \caption{IT Monitoring (3Q, 5C)}
         \label{fig:itmonitoring}
     \end{subfigure}
     \hfill
     \begin{subfigure}[b]{0.27\textwidth}
         \centering
         \includegraphics[width=\textwidth]{figures/customer_service.png}
          \caption{Customer Service (10Q, 6C)}
         \label{fig:customerservice}
     \end{subfigure} 
    \vspace{-3mm}
     \caption{\vspace{-2mm}Six real-world dashboards sourced from Tableau Public (links provided in supplemental materials).
     }
     \label{fig:demo_dashboards}
\end{figure*}

For a data exploration benchmark to be useful to the community, it must provide new workloads  (i.e., exploration scenarios) for DBMSs. In this section, we demonstrate the value of our benchmark by comparing it to an existing benchmark, IDEBench~\cite{idebench}.
Note that since other visualization benchmarks (e.g., \cite{ghita2020white,crossfilter}) are unable to generate simulations with new datasets, we omit them in our experiments.


\begin{table}[]
\caption{Experiments were organized by three parameters: input dataset size, the sequence of goal templates driving the simulations (workflows), and the dashboards tested.}
\vspace{-2mm}
\begin{tabular}{|l|l|l|}
\hline
\textbf{Dataset Size}                                                  & \textbf{Goal Sequence}                                                               & \textbf{Dashboard}                                                                                                                       \\ \hline
\begin{tabular}[c]{@{}l@{}}100K Rows\\ 1M Rows\\ 10M Rows\end{tabular} & \begin{tabular}[c]{@{}l@{}}Shneiderman\\ Battle \& Heer\\ Battle et al.\end{tabular} & \begin{tabular}[c]{@{}l@{}}UBC Energy Map\\ Circulation Activity\\ Customer Service\\ IT Monitor\\ MyRide\\ Supply Chain\end{tabular} \\ \hline
\end{tabular}
\label{tab:experiment-parameters}
\end{table}

\subsection{Identifying Relevant Dashboards and Goals}
\label{sec:goals:examples}

To test the SIMBA benchmark, we selected six representative dashboards diversified across our templates in \autoref{sec:goals:templates} and the
dashboard types classified by Sarikaya et al. ~\cite{sarikaya2019} (see \autoref{fig:demo_dashboards}):

\begin{description}
    \item{\textbf{Circulation Activity by Library} (Strategic decision making): This dashboard provides a strategic analysis of circulation events, both system wide and for individual branches. Users can customize the display by date range and Library branch.
    }
    \item{\textbf{Supply Chain} (Strategic decision making): Focused on the strategic evaluation of order logistics, it allows users to delve into data regarding products, shipping duration, modes, and costs. It also includes filters for regional and categorical analysis.}
    \item{\textbf{UBC Energy Map} (Strategic decision making): it aggregates energy usage data for the UBC campus buildings, offering granular details on energy types and usage per selected building.}
    \item{\textbf{MyRide} (Quantified Self): it visualizes heart rate data throughout an individual's cycling route in Orlando, FL.
    }
    \item{\textbf{IT Monitor} (Operational decision making): Designed for IT professionals, it displays system telemetry and supports in-depth examination of anomalies.
    }
    \item{\textbf{Customer Service} (Operational Decision-Making): it summarizes call center metrics, including performance by individual representatives and call queues, with the capability to filter and highlight specific data for detailed operational analysis.}
\end{description}
We provide links to the dashboards in our supplemental materials\footnote{\url{https://osf.io/vbm8z/?view_only=2e06892f0c104a9e911e8e7599deb2ab}}.



\subsection{Experiment Setup}
\label{sec:evaluation:setup}

\subsubsection{Environment Setup}

Each experiment was conducted on a single server (Linux 5.14.0) with 503GB of memory, 48 cores (Intel(R) Xeon(R) CPU E5-2690 v3 @ 2.60GHz), and 5.5TB of disc space.

\subsubsection{DBMSs and Drivers} Consistent with prior work~\cite{idebench,crossfilter}, we tested our benchmark against multiple data processing back-ends, using the default settings for each. We implemented DBMS wrappers for four systems: PostgreSQL, DuckDB, SQLite, and MonetDB,  which were selected for their support for analytics, usage in interactive applications, and popularity. Datasets were denormalized and no indexing or caching was applied.

\begin{figure*}
    \centering
    \includegraphics[width=0.9\textwidth]{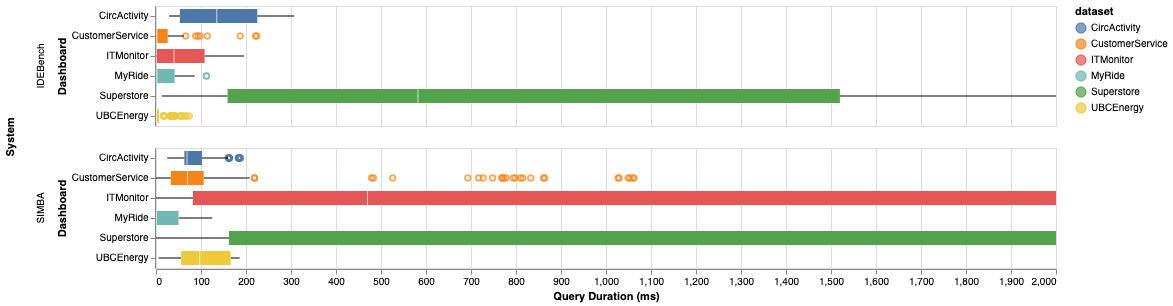}
    \vspace{-3mm}
    \caption{DuckDB's performance for each dashboard tested (10M rows only, figure clipped at 2000ms for sake of space).\vspace{-2mm}
    }
    \label{fig:total_performance}
\end{figure*}

\subsubsection{Benchmark Parameters}
We permute three parameters in our experiments (see \autoref{tab:experiment-parameters}): dataset size, goal template sequences, and dashboards.
 We conducted our experiments using all six of the dashboards in \autoref{fig:demo_dashboards}.
Each dashboard visualizes a different dataset with
different quantitative (Q), and categorical (C) data columns.
To see how performance changes as dataset size increases, we test three different dataset sizes (adopting dataset generation techniques from prior work~\cite{crossfilter,idebench}): 100K, 1M, and 10M rows.

We provide three ordered sequences of goal templates, i.e., benchmark \emph{workflows}, to guide the benchmark simulations (see \autoref{sec:simulation:targeted} for details). These workflows are derived directly from the data exploration goals proposed by Shneiderman~\cite{shneiderman96eyes}, Battle \& Heer~\cite{Bat19}, and the developers of the Crossfilter benchmark~\cite{crossfilter}. These workflow enable our benchmark to re-create established data exploration scenarios with new datasets and dashboards. Note that the aforementioned datasets may contain more data columns than are used in the dashboards, and one dashboard (MyRide) was not compatible with all three workflows. Specifically, the MyRide dashboard contains a low number of quantitative data columns for testing correlations, making it inapplicable to the Battle \& Heer~\cite{Bat19} and crossfilter~\cite{crossfilter} workflows. For each parameter combination, we complete 8 runs with these parameters for each DBMS tested.

\subsubsection{IDEBench Setup.} Only two of our three parameters apply for IDEBench (dataset size, dashboards), since it does not support simulation of exploration goals. We used a mixed workflow, and the default probabilities for generating actions in IDEBench.

\subsubsection{Evaluation Metrics} We focus on the common metric of \emph{query duration}~\cite{idebench,crossfilter,battle2016dynamic}, i.e., the average latency of all queries issued for a given dashboard. Specifically, we observe how query durations increase or decrease across dashboards, dataset sizes, and workflows. The SIMBA benchmark also supports alternative metrics such as response rate~\cite{crossfilter}. However, response rate thresholds must be tailored to the specific requirements of the target dashboard(s)~\cite{crossfilter,Liu14}. Given we are evaluating six different dashboards with distinct response rate characteristics, we omit this metric to save space.



\subsection{SIMBA Workload Analysis}
\label{sec:evaluation:exp1}
We generated workflows for each goal sequence and dashboard combination, then executed each one against various dataset sizes in each DBMS. We evaluate the differences in the workflows themselves as well as the output performance metrics.

\paragraph{Differences in Dashboards Lead to Differences in DBMS Performance}

We observed wide variations in performance across all dashboards as measured by average query duration (see \autoref{fig:total_performance}). The My Ride, Customer Service and Circulation Activity dashboards all achieved an average query duration less than 100ms, with a relatively small amount of variability across queries. This contrasts with the Superstore, IT Monitor and UBC Energy dashboards which not only reported longer average query duration (3,145ms, 741ms, and 243ms), but also a larger inter-quartile range. These variations can be explained in part by the complexity of the underlying dataset, but also the specific data columns used in a visualization and aggregation type and the interactive components.

The Circulation Activity and My Ride dashboards contain only two visualizations, resulting in workflows that contain only two aggregated queries with various WHERE clauses depending on the filters that were applied on the visualization. In the case of Circulation Activity, the visualizations and supporting queries are identical, hence the lack of variability in query duration. 

Customer Service is a more complicated dashboard with five visualizations that filter each other and four interactive components in total. Examining query performance by visualization we note that three of the visualizations achieve similar durations of 736-1,123ms, while the Calls per Rep and Total Calls per Hour visualizations are 3-5x slower (3,518ms, 5,386ms), likely due to requesting more data columns.  Therefore, the construction of visualizations in the dashboard is relevant to performance.

\begin{figure*}
    \centering
    \includegraphics[width=0.9\textwidth]{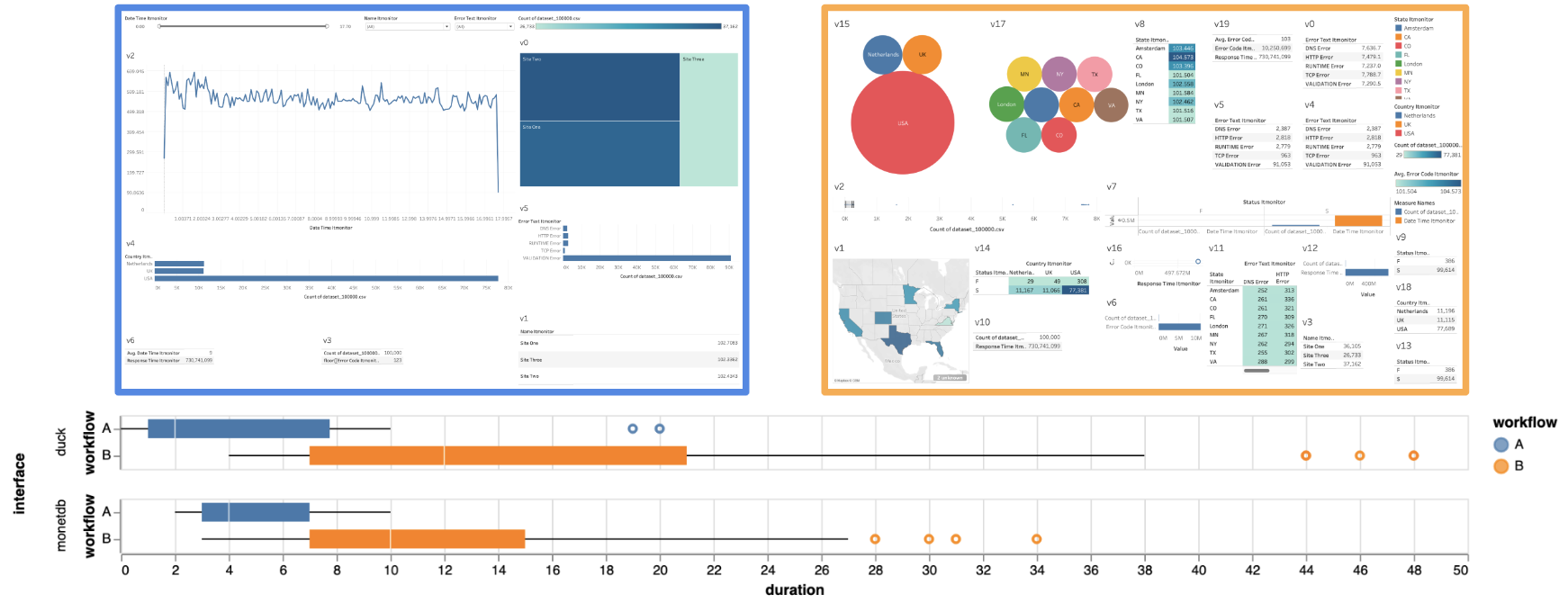}
    \vspace{-3mm}
    \caption{Hand-tuned dashboards matching two workflows generated by IDEBench and their corresponding performance numbers. Not only do the dashboards differ in visualization complexity, both data column usage and visualization numbers, from the original, but these design differences also yield varying performance. }
    \label{fig:itmonitor_idebench_combined}
    \vspace{-3mm}
\end{figure*}

As noted in \autoref{sec:comparison}, SIMBA workflows are constrained by a dashboard specification while IDEBench's are not. To demonstrate the significance of this constraint, we generated 50 IDEBench workflows for the IT Monitor dataset and reverse engineered dashboard configurations. \autoref{fig:itmonitor_idebench_combined} shows two (stylized) dashboards generated by this process that vary substantially in both the number of visualizations and data column usage across visualizations. Across the 50 workflows, IDEBench created an average of 13 visualizations (min=7, max=20), substantially more complex than the original IT Monitor dashboard which contains 3. The visualizations are also densely linked; on average a single interaction triggers 9 visualization updates (avg=9, min=1, max=15).

Given our findings, high variability in dashboard design may lead to wild fluctuations in perceived DBMS performance. While some benchmark variability is advantageous to enable users to simulate unique runs, \textbf{too much variability can lead to the generation of unrealistic dashboards} that may not be meaningful to test against. With significant variation between IDEBench runs, it is unclear whether observed differences in DBMS performance are due to randomness or significant differences in dashboard design.

%

\begin{table*}[]

\caption{Average and standard deviations for the SIMBA workloads for two dashboards.}
\vspace{-3mm}
\begin{tabular}{c|c|c|c}
\toprule
\textbf{Statistic}& \textbf{Count Categorical and Quantitative Data Columns}&  \textbf{Count Aggregated Data Columns}&\textbf{Count Filters}\\ \midrule
\textbf{Customer Service}& $1.5 \pm1.3$&  $1.0 \pm0$&$1.9 \pm0.9$\\ 
 \textbf{IT Monitor}&$3.0 \pm1.2$& $0.8 \pm2.0$&$5.8 \pm0.8$\\ \bottomrule
\end{tabular}
\label{tab:perform}
\vspace{-2mm}

\end{table*}

\begin{figure}
    \centering
    \includegraphics[width=0.9\columnwidth]{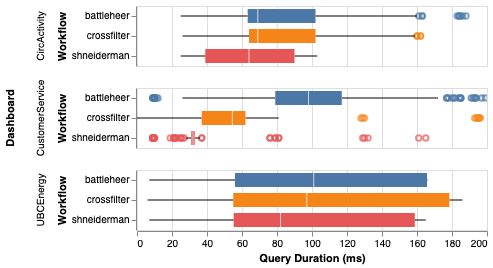}
    \vspace{-3mm}
    \caption{Distribution of query durations grouped by workflow and dashboard.}
    \label{fig:exp1-workflow}
    \vspace{-3mm}
\end{figure}

\paragraph{Differences in Workflows Influence DBMS Performance.} In addition to the design of the dashboard, we find that workflow differences can also impact the performance of the DBMS ( \autoref{fig:exp1-workflow} ). Regardless of the dashboard, the Shneiderman workflow achieved the lowest query duration, although in some cases, performance did not change substantially across workflows, such as for the Circulation Activity dashboard. These cases tended to occur for datasets with few data attributes and dashboards with similar visualizations. For example, the Circulation Activity dashboard contains only two visualizations, leaving limited options for variation in SQL queries. However, the Customer Service dashboard achieves significantly different performance depending on the workflow.

As discussed earlier in this section, the dashboards created by IDEBench workflows for the IT Monitor dataset are dense with visualizations and links. Although there were more total visualizations in each dashboard, each visualization was simpler than those in SIMBA's dashboards. IDEBench workflows have an average of 2.1 data attributes and 13.2 filters per visualization compared to 3.8 and 5.8 in SIMBA (\autoref{tab:perform}). These findings suggest that \textbf{SIMBA balances both visualization and filtering complexity when generating query workloads, whereas IDEBench emphasizes adding filters.} A single interaction in IDEBench filters numerous visualizations, in turn triggering many simultaneous queries. SIMBA generates fewer, but also more complex queries.



\subsection{User Study}
\label{sec:evaluation:user-study}


We conducted a user study
to answer two questions: (1) how realistic are the SIMBA simulations? and (2) how useful do experts find the benchmark? This study was approved by our institution's IRB.

\paragraph{Study Design} Our study was conducted in two parts. First, we recruited six analysts (three from a university, three from a company) to explore data using two existing dashboards
(IT Monitoring, Customer Service). We recreated the dashboards in Mosaic~\cite{heer2024mosaic} so that we could log analysts' interactions and the corresponding SQL queries. Analysts spent 12 minutes exploring each dashboard, consistent with previous studies on exploratory data analysis ~\cite{Bat19,zgraggen2016progressive,zgraggen2018investigating,crossfilter}.
Analysts completed the study in 30 minutes. This process generated more data than we needed, so we randomly selected a subset of logs for part two.

Next, we recruited six database and visualization experts (three experts from different companies/government organizations, three PhD students from the same university) to compare the analysts' logs against separate logs generated by the SIMBA benchmark.
We generated SIMBA logs to match the average length of the human analysts' logs and used the same SIMBA randomization settings for both dashboards. IDEBench was omitted from the study because it does not support pre-specified dashboards in its simulations. The logs were provided in a Google spreadsheet along with links to the dashboards, so that the experts could compare the dashboards with the log entries.
{We asked the experts to guess which logs were generated by SIMBA.} The experts spent 15 minutes comparing the pair of logs for each of the two dashboards.
Finally, we gave the experts an overview of the SIMBA benchmark and solicited feedback on its usability/utility. Experts spent 50-60 minutes completing the study.

\paragraph{Log Comparison Results} 
If SIMBA's logs are sufficiently realistic then we would expect the experts' guesses to be no better than random chance. \textbf{Fifty percent of the time (6/12 total guesses) the experts correctly identified SIMBA's logs.} We performed a binomial test comparing experts' guesses to a null hypothesis of success rate $\leq$50\%, and found that the probability of 7 or more successes is 38.7\% (p-value=.774). 


%

When we asked about their log analysis strategies, experts pointed out that for the IT Monitoring dashboard (5/6 expert successes), SIMBA's logs repeatedly emit SQL queries returning zero results. Human analysts occasionally performed interactions that resulted in empty visualizations, but would rarely repeat this error in the same session. We note that SIMBA's Oracle model would not produce this error either, since none of the specified goal queries produce empty query results. Thus, this error results when the open-ended (i.e., Markov) model is used. In other words, the randomization level was too high for the IT monitoring dashboard. These findings suggest that \textbf{SIMBA's randomization parameters are sensitive to the underlying dashboard design. For a simpler dashboard like Customer Service (1/6 expert successes), higher randomization is acceptable and will not cause unrealistic errors (i.e., too many empty query results). For dashboards with more filters like IT Monitoring, adding too many random filters may produce unrealistic results, requiring lower randomization.}

\paragraph{Expert Feedback} All of the experts mentioned that SIMBA could support their work (e.g., for anticipating dashboard user behavior, performance testing, or dashboard optimization), but experts also expressed caution in utilizing it effectively (experts labeled as E1, E2, etc.). Two experts mentioned that SIMBA could be useful for performance testing. E2 said ``\emph{If you're able to get a bunch of variety and then work out some kinks or things that were unexpected before having to hire a bunch of analysts or anything like that, then I can definitely see the utility.'' Similarly, E4 said: ``\emph{I would definitely use it as a starting point to make sure that my [dashboard] works and it doesn't cause a catastrophic failure when certain queries run.}'' However, E2 also stressed that SIMBA logs should be considered in aggregate only: ``\emph{Where I would be cautious is in assuming that you can run one test and that's the one test of how everyone will be interacting, because that's obviously not the case}.'' E1 posited that SIMBA could help database administrators reason about future dashboard usage: ``\emph{you generally go to a dashboard trying to answer a question, you know, so trying to simulate those questions would definitely give the DBAs a better idea of... how you might answer questions from your dashboard.}''
E5 said that adding indexes later can be difficult when datasets grow to billions of records, and ``\emph{if you could mock that ahead of time, understand user behavior---a web app for a dashboard---is similar type of behavior, I think it can be beneficial for sure.}'' However, E3 considers it ``risky'' to use simulations to direct optimization: ``\emph{I would feel much safer... letting people use the unoptimized version [of the dashboard] and then optimize it.}''}

\subsection{Evaluation Takeaways}

Our analysis in \autoref{sec:evaluation:exp1} shows that dashboard design significantly impacts query diversity in simulated workloads, demonstrating the importance of supporting dashboard customization in DBMS performance benchmarks for visual analysis scenarios. By considering these aspects, SIMBA provides more realistic performance evaluations compared to existing benchmarks like IDEbench \cite{idebench} which may generate unrealistic workloads due to high variability.

Our user study in \autoref{sec:evaluation:user-study} confirms that SIMBA can generate realistic user interaction logs that are largely indistinguishable from actual user behavior. Expert feedback suggests that while SIMBA can be useful for anticipating user behavior and performance testing, randomization parameters must be tuned carefully to avoid unrealistic simulations, especially in more complex dashboards.
\section{Related Work}

Here, we summarize relevant research for benchmarking system performance in exploratory visual analysis (or EVA~\cite{Bat19}) scenarios.

\paragraph{Mapping Visualizations and Interactions to SQL} The SIMBA benchmark extends prior work in mapping visualization interfaces to SQL~\cite{siddiqui2016effortless,stolte2002polaris,yan2021tessera,zhang2019mining}.
For example, Stolte et al. propose a SQL-based table algebra for specifying complex, grouped aggregation visualizations~\cite{stolte2002polaris},
enabling Polaris (later Tableau) to map visualizations to SQL queries for efficient processing.
However, the SIMBA benchmark supports a wider range of visualization interface designs compared to these approaches. Chen and Wu use a similar idea to generate dashboard designs from a set of target queries~\cite{chen2022pi2} but do not simulate how users might interact with these dashboards. The SIMBA benchmark is the \emph{first} DBMS benchmark to apply this approach to simulate realistic exploration behaviors.

\paragraph{Measuring EVA Performance} Many metrics have been proposed in the literature to gauge exploration performance. In this paper, we focus on metrics that can be calculated directly from user interaction logs.
Some metrics assess the \emph{pacing and flow} of interactions over time, including interaction rates~\cite{Liu14, feng2019patterns, zgraggen2016progressive}. A related measure frequently used is system response time~\cite{Liu14,zgraggen2016progressive,idreos2015overview,jiang2018evaluating}. Average or worst case latency per interaction is also considered~\cite{battle2016dynamic, chan2008maintaining,crotty2016case,kamat2014distributed, rahman2017seen}.
Some also consider summative measures such as total exploration time~\cite{dimitriadou2014explore, feng2019patterns}, total interactions performed~\cite{dabek2017grammar,dimitriadou2014explore, feng2019patterns, gotz2009behavior}, or attributes explored~\cite{wongsuphasawat2017voyager}. The SIMBA benchmark easily supports computing all of these measures, as well as new measures, such as the measures of interaction variance we discuss in \autoref{sec:evaluation}.

\paragraph{Benchmarking EVA Tools} Many benchmarks have been developed in the database and visualization communities.  However, the vast majority fail to fully capture the key characteristics and concerns of data exploration~\cite{battle2017benchmark,crossfilter}. For example,
DBMS benchmarks such as TPC-H, TPC-DS, and SSB simulate how well a DBMS supports a range of routine business intelligence operations~\cite{barata2015overview,o2007star}; however, analysts often query their data opportunistically based on statistical patterns or anomalies seen as they explore, resulting in spontaneous queries that these benchmarks fail to capture~\cite{crossfilter}.
Visualization benchmarks such as the Visual Analytics Benchmark Repository~\cite{plaisant2007promoting} record realistic exploration use cases, but act more as unstructured data archives than true benchmarks~\cite{battle2017benchmark}.
A few benchmarks exist for interactive scenarios~\cite{idebench,rahman2020benchmarking,crossfilter}, however they focus on only one or two basic use cases, e.g., only bar chart exploration~\cite{crossfilter,idebench}. Little has been done to make these benchmarks more robust, relevant, and easy to adopt across multiple communities. Our SIMBA benchmark approach addresses these problems.

\section{Conclusion}
In this paper, we presented a SIMulation-BAsed (SIMBA) benchmark for evaluating database management systems connected to any standard data exploration dashboard. 
SIMBA models user analysis goals as a query set generated through a sequence of user interactions. We compared the flexibility of SIMBA against IDEBench using four DBMSs connected to six unique dashboards.
We found that differences in dashboard designs induced measurable differences in exploration patterns and ultimately DBMS performance, even when the user goals were the same. Similarly, different goals lead to variation in DBMS performance, even with the same dashboard. SIMBA successfully simulated all of these performance variations whereas IDEBench only supported a subset of them, showcasing the value in controlling for interface design and user analysis goals alongside dataset schema, dataset size, and simulated interactions.
\begin{acks}
This research was supported in part by the NSF (award \# 2141506 and CSGrad4US award \# 2313998) and Google. We thank the Data Systems Lab at Maryland, the UW Database Group, the UW IDL, and our reviewers for their invaluable feedback.
\end{acks}

\bibliographystyle{ACM-Reference-Format}
\bibliography{main}


\begin{thebibliography}{49}


\ifx \showCODEN    \undefined \def \showCODEN     #1{\unskip}     \fi
\ifx \showDOI      \undefined \def \showDOI       #1{#1}\fi
\ifx \showISBNx    \undefined \def \showISBNx     #1{\unskip}     \fi
\ifx \showISBNxiii \undefined \def \showISBNxiii  #1{\unskip}     \fi
\ifx \showISSN     \undefined \def \showISSN      #1{\unskip}     \fi
\ifx \showLCCN     \undefined \def \showLCCN      #1{\unskip}     \fi
\ifx \shownote     \undefined \def \shownote      #1{#1}          \fi
\ifx \showarticletitle \undefined \def \showarticletitle #1{#1}   \fi
\ifx \showURL      \undefined \def \showURL       {\relax}        \fi
\providecommand\bibfield[2]{#2}
\providecommand\bibinfo[2]{#2}
\providecommand\natexlab[1]{#1}
\providecommand\showeprint[2][]{arXiv:#2}

\bibitem[\protect\citeauthoryear{Adobe}{Adobe}{2022}]%
        {adobe2022panels}
\bibfield{author}{\bibinfo{person}{Adobe}.} \bibinfo{year}{2022}\natexlab{}.
\newblock \bibinfo{title}{{Panels Overview | Adobe Analytics}}.
\newblock \bibinfo{howpublished}{{\url{https://experienceleague.adobe.com/docs/analytics/analyze/analysis-workspace/panels/panels.html?lang=en}}}.
\newblock


\bibitem[\protect\citeauthoryear{Alspaugh, Zokaei, Liu, Jin, and Hearst}{Alspaugh et~al\mbox{.}}{2019}]%
        {alspaugh2019futzing}
\bibfield{author}{\bibinfo{person}{Sara Alspaugh}, \bibinfo{person}{Nava Zokaei}, \bibinfo{person}{Andrea Liu}, \bibinfo{person}{Cindy Jin}, {and} \bibinfo{person}{Marti~A. Hearst}.} \bibinfo{year}{2019}\natexlab{}.
\newblock \showarticletitle{Futzing and Moseying: Interviews with Professional Data Analysts on Exploration Practices}.
\newblock \bibinfo{journal}{\emph{IEEE TVCG}} \bibinfo{volume}{25}, \bibinfo{number}{1} (\bibinfo{year}{2019}), \bibinfo{pages}{22--31}.
\newblock
\urldef\tempurl%
\url{https://doi.org/10.1109/TVCG.2018.2865040}
\showDOI{\tempurl}


\bibitem[\protect\citeauthoryear{Bach, Freeman, Abdul-Rahman, Turkay, Khan, Fan, and Chen}{Bach et~al\mbox{.}}{2022}]%
        {bach2022dashboard}
\bibfield{author}{\bibinfo{person}{Benjamin Bach}, \bibinfo{person}{Euan Freeman}, \bibinfo{person}{Alfie Abdul-Rahman}, \bibinfo{person}{Cagatay Turkay}, \bibinfo{person}{Saiful Khan}, \bibinfo{person}{Yulei Fan}, {and} \bibinfo{person}{Min Chen}.} \bibinfo{year}{2022}\natexlab{}.
\newblock \showarticletitle{Dashboard Design Patterns}.
\newblock \bibinfo{journal}{\emph{IEEE TVCG}} (\bibinfo{year}{2022}), \bibinfo{pages}{1--11}.
\newblock
\urldef\tempurl%
\url{https://doi.org/10.1109/TVCG.2022.3209448}
\showDOI{\tempurl}


\bibitem[\protect\citeauthoryear{Barata, Bernardino, and Furtado}{Barata et~al\mbox{.}}{2015}]%
        {barata2015overview}
\bibfield{author}{\bibinfo{person}{Melyssa Barata}, \bibinfo{person}{Jorge Bernardino}, {and} \bibinfo{person}{Pedro Furtado}.} \bibinfo{year}{2015}\natexlab{}.
\newblock \showarticletitle{An overview of decision support benchmarks: TPC-DS, TPC-H and SSB}.
\newblock In \bibinfo{booktitle}{\emph{New Contributions in Information Systems and Technologies}}. \bibinfo{publisher}{Springer}, \bibinfo{pages}{619--628}.
\newblock


\bibitem[\protect\citeauthoryear{Battle, Angelini, Binnig, Catarci, Eichmann, Fekete, Santucci, Sedlmair, and Willett}{Battle et~al\mbox{.}}{2018}]%
        {battle2018evaluating}
\bibfield{author}{\bibinfo{person}{Leilani Battle}, \bibinfo{person}{Marco Angelini}, \bibinfo{person}{Carsten Binnig}, \bibinfo{person}{Tiziana Catarci}, \bibinfo{person}{Philipp Eichmann}, \bibinfo{person}{Jean-Daniel Fekete}, \bibinfo{person}{Giuseppe Santucci}, \bibinfo{person}{Michael Sedlmair}, {and} \bibinfo{person}{Wesley Willett}.} \bibinfo{year}{2018}\natexlab{}.
\newblock \showarticletitle{Evaluating Visual Data Analysis Systems: A Discussion Report}. In \bibinfo{booktitle}{\emph{Proceedings of the Workshop on Human-In-the-Loop Data Analytics}} (Houston, TX, USA) \emph{(\bibinfo{series}{HILDA '18})}. \bibinfo{publisher}{Association for Computing Machinery}, \bibinfo{address}{New York, NY, USA}, Article \bibinfo{articleno}{4}, \bibinfo{numpages}{6}~pages.
\newblock
\showISBNx{9781450358279}
\urldef\tempurl%
\url{https://doi.org/10.1145/3209900.3209901}
\showDOI{\tempurl}


\bibitem[\protect\citeauthoryear{Battle, Chang, Heer, and Stonebraker}{Battle et~al\mbox{.}}{2017}]%
        {battle2017benchmark}
\bibfield{author}{\bibinfo{person}{L. Battle}, \bibinfo{person}{R. Chang}, \bibinfo{person}{J. Heer}, {and} \bibinfo{person}{M. Stonebraker}.} \bibinfo{year}{2017}\natexlab{}.
\newblock \showarticletitle{Position statement: The case for a visualization performance benchmark}. In \bibinfo{booktitle}{\emph{2017 IEEE Workshop on Data Systems for Interactive Analysis (DSIA)}}. \bibinfo{pages}{1--5}.
\newblock
\urldef\tempurl%
\url{https://doi.org/10.1109/DSIA.2017.8339089}
\showDOI{\tempurl}


\bibitem[\protect\citeauthoryear{Battle, Chang, and Stonebraker}{Battle et~al\mbox{.}}{2016}]%
        {battle2016dynamic}
\bibfield{author}{\bibinfo{person}{Leilani Battle}, \bibinfo{person}{Remco Chang}, {and} \bibinfo{person}{Michael Stonebraker}.} \bibinfo{year}{2016}\natexlab{}.
\newblock \showarticletitle{Dynamic prefetching of data tiles for interactive visualization}. In \bibinfo{booktitle}{\emph{Proceedings of the 2016 International Conference on Management of Data}}. \bibinfo{pages}{1363--1375}.
\newblock


\bibitem[\protect\citeauthoryear{Battle, Eichmann, Angelini, Catarci, Santucci, Zheng, Binnig, Fekete, and Moritz}{Battle et~al\mbox{.}}{2020}]%
        {crossfilter}
\bibfield{author}{\bibinfo{person}{Leilani Battle}, \bibinfo{person}{Philipp Eichmann}, \bibinfo{person}{Marco Angelini}, \bibinfo{person}{Tiziana Catarci}, \bibinfo{person}{Giuseppe Santucci}, \bibinfo{person}{Yukun Zheng}, \bibinfo{person}{Carsten Binnig}, \bibinfo{person}{Jean-Daniel Fekete}, {and} \bibinfo{person}{Dominik Moritz}.} \bibinfo{year}{2020}\natexlab{}.
\newblock \showarticletitle{Database Benchmarking for Supporting Real-Time Interactive Querying of Large Data}. In \bibinfo{booktitle}{\emph{Proceedings of the 2020 ACM SIGMOD International Conference on Management of Data}} (Portland, OR, USA) \emph{(\bibinfo{series}{SIGMOD ’20})}. \bibinfo{publisher}{Association for Computing Machinery}, \bibinfo{address}{New York, NY, USA}, \bibinfo{pages}{1571–1587}.
\newblock
\showISBNx{9781450367356}
\urldef\tempurl%
\url{https://doi.org/10.1145/3318464.3389732}
\showDOI{\tempurl}


\bibitem[\protect\citeauthoryear{Chan, Xiao, Gerth, and Hanrahan}{Chan et~al\mbox{.}}{2008}]%
        {chan2008maintaining}
\bibfield{author}{\bibinfo{person}{Sye-Min Chan}, \bibinfo{person}{Ling Xiao}, \bibinfo{person}{John Gerth}, {and} \bibinfo{person}{Pat Hanrahan}.} \bibinfo{year}{2008}\natexlab{}.
\newblock \showarticletitle{Maintaining interactivity while exploring massive time series}. In \bibinfo{booktitle}{\emph{2008 IEEE Symposium on Visual Analytics Science and Technology}}. \bibinfo{pages}{59--66}.
\newblock
\urldef\tempurl%
\url{https://doi.org/10.1109/VAST.2008.4677357}
\showDOI{\tempurl}


\bibitem[\protect\citeauthoryear{Chen and Wu}{Chen and Wu}{2022}]%
        {chen2022pi2}
\bibfield{author}{\bibinfo{person}{Yiru Chen} {and} \bibinfo{person}{Eugene Wu}.} \bibinfo{year}{2022}\natexlab{}.
\newblock \showarticletitle{PI2: End-to-End Interactive Visualization Interface Generation from Queries}. In \bibinfo{booktitle}{\emph{Proceedings of the 2022 International Conference on Management of Data}} (Philadelphia, PA, USA) \emph{(\bibinfo{series}{SIGMOD '22})}. \bibinfo{publisher}{Association for Computing Machinery}, \bibinfo{address}{New York, NY, USA}, \bibinfo{pages}{1711–1725}.
\newblock
\showISBNx{9781450392495}
\urldef\tempurl%
\url{https://doi.org/10.1145/3514221.3526166}
\showDOI{\tempurl}


\bibitem[\protect\citeauthoryear{Crotty, Galakatos, Zgraggen, Binnig, and Kraska}{Crotty et~al\mbox{.}}{2016}]%
        {crotty2016case}
\bibfield{author}{\bibinfo{person}{Andrew Crotty}, \bibinfo{person}{Alex Galakatos}, \bibinfo{person}{Emanuel Zgraggen}, \bibinfo{person}{Carsten Binnig}, {and} \bibinfo{person}{Tim Kraska}.} \bibinfo{year}{2016}\natexlab{}.
\newblock \showarticletitle{The case for interactive data exploration accelerators (IDEAs)}. In \bibinfo{booktitle}{\emph{Proceedings of the Workshop on Human-In-the-Loop Data Analytics}}. \bibinfo{pages}{1--6}.
\newblock


\bibitem[\protect\citeauthoryear{Dabek and Caban}{Dabek and Caban}{2017}]%
        {dabek2017grammar}
\bibfield{author}{\bibinfo{person}{Filip Dabek} {and} \bibinfo{person}{Jesus~J Caban}.} \bibinfo{year}{2017}\natexlab{}.
\newblock \showarticletitle{A Grammar-based Approach for Modeling User Interactions and Generating Suggestions During the Data Exploration Process}.
\newblock \bibinfo{journal}{\emph{IEEE TVCG}} \bibinfo{volume}{23}, \bibinfo{number}{1} (\bibinfo{year}{2017}), \bibinfo{pages}{41--50}.
\newblock
\urldef\tempurl%
\url{https://doi.org/10.1109/TVCG.2016.2598471}
\showDOI{\tempurl}


\bibitem[\protect\citeauthoryear{Dimitriadou, Papaemmanouil, and Diao}{Dimitriadou et~al\mbox{.}}{2014}]%
        {dimitriadou2014explore}
\bibfield{author}{\bibinfo{person}{Kyriaki Dimitriadou}, \bibinfo{person}{Olga Papaemmanouil}, {and} \bibinfo{person}{Yanlei Diao}.} \bibinfo{year}{2014}\natexlab{}.
\newblock \showarticletitle{Explore-by-example: An Automatic Query Steering Framework for Interactive Data Exploration}. In \bibinfo{booktitle}{\emph{Proceedings of the 2014 ACM SIGMOD International Conference on Management of Data}} (Snowbird, Utah, USA) \emph{(\bibinfo{series}{SIGMOD '14})}. \bibinfo{publisher}{ACM}, \bibinfo{address}{New York, NY, USA}, \bibinfo{pages}{517--528}.
\newblock
\showISBNx{978-1-4503-2376-5}
\urldef\tempurl%
\url{https://doi.org/10.1145/2588555.2610523}
\showDOI{\tempurl}


\bibitem[\protect\citeauthoryear{Eichmann, Zgraggen, Binnig, and Kraska}{Eichmann et~al\mbox{.}}{2020}]%
        {idebench}
\bibfield{author}{\bibinfo{person}{Philipp Eichmann}, \bibinfo{person}{Emanuel Zgraggen}, \bibinfo{person}{Carsten Binnig}, {and} \bibinfo{person}{Tim Kraska}.} \bibinfo{year}{2020}\natexlab{}.
\newblock \showarticletitle{IDEBench: A Benchmark for Interactive Data Exploration}. In \bibinfo{booktitle}{\emph{Proceedings of the 2020 ACM SIGMOD International Conference on Management of Data}} (Portland, OR, USA) \emph{(\bibinfo{series}{SIGMOD ’20})}. \bibinfo{publisher}{Association for Computing Machinery}, \bibinfo{address}{New York, NY, USA}, \bibinfo{pages}{1555–1569}.
\newblock
\showISBNx{9781450367356}
\urldef\tempurl%
\url{https://doi.org/10.1145/3318464.3380574}
\showDOI{\tempurl}


\bibitem[\protect\citeauthoryear{Feng, Peck, and Harrison}{Feng et~al\mbox{.}}{2019}]%
        {feng2019patterns}
\bibfield{author}{\bibinfo{person}{M. Feng}, \bibinfo{person}{E. Peck}, {and} \bibinfo{person}{L. Harrison}.} \bibinfo{year}{2019}\natexlab{}.
\newblock \showarticletitle{Patterns and Pace: Quantifying Diverse Exploration Behavior with Visualizations on the Web}.
\newblock \bibinfo{journal}{\emph{IEEE TVCG}} \bibinfo{volume}{25}, \bibinfo{number}{1} (\bibinfo{date}{Jan} \bibinfo{year}{2019}), \bibinfo{pages}{501--511}.
\newblock
\showISSN{1077-2626}
\urldef\tempurl%
\url{https://doi.org/10.1109/TVCG.2018.2865117}
\showDOI{\tempurl}


\bibitem[\protect\citeauthoryear{Ghallab, Nau, and Traverso}{Ghallab et~al\mbox{.}}{2016}]%
        {ghallab2016automated}
\bibfield{author}{\bibinfo{person}{Malik Ghallab}, \bibinfo{person}{Dana Nau}, {and} \bibinfo{person}{Paolo Traverso}.} \bibinfo{year}{2016}\natexlab{}.
\newblock \bibinfo{booktitle}{\emph{Automated planning and acting}}.
\newblock \bibinfo{publisher}{Cambridge University Press}.
\newblock


\bibitem[\protect\citeauthoryear{Ghita, Tom{\'e}, and Boncz}{Ghita et~al\mbox{.}}{2020}]%
        {ghita2020white}
\bibfield{author}{\bibinfo{person}{Bogdan Ghita}, \bibinfo{person}{Diego~G Tom{\'e}}, {and} \bibinfo{person}{Peter~A Boncz}.} \bibinfo{year}{2020}\natexlab{}.
\newblock \showarticletitle{White-box Compression: Learning and Exploiting Compact Table Representations.}. In \bibinfo{booktitle}{\emph{CIDR}}.
\newblock


\bibitem[\protect\citeauthoryear{Gotz and Wen}{Gotz and Wen}{2009}]%
        {gotz2009behavior}
\bibfield{author}{\bibinfo{person}{David Gotz} {and} \bibinfo{person}{Zhen Wen}.} \bibinfo{year}{2009}\natexlab{}.
\newblock \showarticletitle{Behavior-Driven Visualization Recommendation}. In \bibinfo{booktitle}{\emph{Proceedings of the 14th International Conference on Intelligent User Interfaces}} (Sanibel Island, Florida, USA) \emph{(\bibinfo{series}{IUI '09})}. \bibinfo{publisher}{Association for Computing Machinery}, \bibinfo{address}{New York, NY, USA}, \bibinfo{pages}{315–324}.
\newblock
\showISBNx{9781605581682}
\urldef\tempurl%
\url{https://doi.org/10.1145/1502650.1502695}
\showDOI{\tempurl}


\bibitem[\protect\citeauthoryear{Heer and Moritz}{Heer and Moritz}{2024}]%
        {heer2024mosaic}
\bibfield{author}{\bibinfo{person}{Jeffrey Heer} {and} \bibinfo{person}{Dominik Moritz}.} \bibinfo{year}{2024}\natexlab{}.
\newblock \showarticletitle{Mosaic: An Architecture for Scalable \& Interoperable Data Views}.
\newblock \bibinfo{journal}{\emph{IEEE Transactions on Visualization and Computer Graphics}} \bibinfo{volume}{30}, \bibinfo{number}{1} (\bibinfo{year}{2024}), \bibinfo{pages}{436--446}.
\newblock
\urldef\tempurl%
\url{https://doi.org/10.1109/TVCG.2023.3327189}
\showDOI{\tempurl}


\bibitem[\protect\citeauthoryear{Idreos, Papaemmanouil, and Chaudhuri}{Idreos et~al\mbox{.}}{2015}]%
        {idreos2015overview}
\bibfield{author}{\bibinfo{person}{Stratos Idreos}, \bibinfo{person}{Olga Papaemmanouil}, {and} \bibinfo{person}{Surajit Chaudhuri}.} \bibinfo{year}{2015}\natexlab{}.
\newblock \showarticletitle{Overview of data exploration techniques}. In \bibinfo{booktitle}{\emph{Proceedings of the 2015 ACM SIGMOD International Conference on Management of Data}}. ACM, \bibinfo{pages}{277--281}.
\newblock


\bibitem[\protect\citeauthoryear{Jiang, Rahman, and Nandi}{Jiang et~al\mbox{.}}{2018}]%
        {jiang2018evaluating}
\bibfield{author}{\bibinfo{person}{Lilong Jiang}, \bibinfo{person}{Protiva Rahman}, {and} \bibinfo{person}{Arnab Nandi}.} \bibinfo{year}{2018}\natexlab{}.
\newblock \showarticletitle{Evaluating interactive data systems: Workloads, metrics, and guidelines}. In \bibinfo{booktitle}{\emph{Proceedings of the 2018 International Conference on Management of Data}}. ACM, \bibinfo{pages}{1637--1644}.
\newblock


\bibitem[\protect\citeauthoryear{Kamat, Jayachandran, Tunga, and Nandi}{Kamat et~al\mbox{.}}{2014}]%
        {kamat2014distributed}
\bibfield{author}{\bibinfo{person}{Niranjan Kamat}, \bibinfo{person}{Prasanth Jayachandran}, \bibinfo{person}{Karthik Tunga}, {and} \bibinfo{person}{Arnab Nandi}.} \bibinfo{year}{2014}\natexlab{}.
\newblock \showarticletitle{Distributed and interactive cube exploration}. In \bibinfo{booktitle}{\emph{2014 IEEE 30th International Conference on Data Engineering}}. \bibinfo{pages}{472--483}.
\newblock
\urldef\tempurl%
\url{https://doi.org/10.1109/ICDE.2014.6816674}
\showDOI{\tempurl}


\bibitem[\protect\citeauthoryear{Lam, Tory, and Munzner}{Lam et~al\mbox{.}}{2018}]%
        {lam2018bridging}
\bibfield{author}{\bibinfo{person}{Heidi Lam}, \bibinfo{person}{Melanie Tory}, {and} \bibinfo{person}{Tamara Munzner}.} \bibinfo{year}{2018}\natexlab{}.
\newblock \showarticletitle{Bridging from Goals to Tasks with Design Study Analysis Reports}.
\newblock \bibinfo{journal}{\emph{IEEE TVCG}} \bibinfo{volume}{24}, \bibinfo{number}{1} (\bibinfo{year}{2018}), \bibinfo{pages}{435--445}.
\newblock
\urldef\tempurl%
\url{https://doi.org/10.1109/TVCG.2017.2744319}
\showDOI{\tempurl}


\bibitem[\protect\citeauthoryear{{Leilani Battle} and {Jeffrey Heer}}{{Leilani Battle} and {Jeffrey Heer}}{2019}]%
        {Bat19}
\bibfield{author}{\bibinfo{person}{{Leilani Battle}} {and} \bibinfo{person}{{Jeffrey Heer}}.} \bibinfo{year}{2019}\natexlab{}.
\newblock \showarticletitle{Characterizing Exploratory Visual Analysis: A Literature Review and Evaluation of Analytic Provenance in Tableau}.
\newblock \bibinfo{journal}{\emph{Computer Graphics Forum}}  \bibinfo{volume}{38} (\bibinfo{date}{06} \bibinfo{year}{2019}), \bibinfo{pages}{145--159}.
\newblock
\urldef\tempurl%
\url{https://doi.org/10.1111/cgf.13678}
\showDOI{\tempurl}


\bibitem[\protect\citeauthoryear{{Matthew Brehmer} and {Tamara Munzner}}{{Matthew Brehmer} and {Tamara Munzner}}{2013}]%
        {Brehmer13}
\bibfield{author}{\bibinfo{person}{{Matthew Brehmer}} {and} \bibinfo{person}{{Tamara Munzner}}.} \bibinfo{year}{2013}\natexlab{}.
\newblock \showarticletitle{A Multi-Level Typology of Abstract Visualization Tasks}.
\newblock \bibinfo{journal}{\emph{IEEE TVCG}} \bibinfo{volume}{19}, \bibinfo{number}{12} (\bibinfo{year}{2013}), \bibinfo{pages}{2376--2385}.
\newblock


\bibitem[\protect\citeauthoryear{Microsoft}{Microsoft}{2022}]%
        {microsoft2022powerbi}
\bibfield{author}{\bibinfo{person}{Microsoft}.} \bibinfo{year}{2022}\natexlab{}.
\newblock \bibinfo{title}{{Data Visualization | Microsoft Power BI}}.
\newblock \bibinfo{howpublished}{{\url{https://powerbi.microsoft.com/en-us/}}}.
\newblock


\bibitem[\protect\citeauthoryear{Moritz, Howe, and Heer}{Moritz et~al\mbox{.}}{2019}]%
        {moritz2019falcon}
\bibfield{author}{\bibinfo{person}{Dominik Moritz}, \bibinfo{person}{Bill Howe}, {and} \bibinfo{person}{Jeffrey Heer}.} \bibinfo{year}{2019}\natexlab{}.
\newblock \showarticletitle{Falcon: Balancing interactive latency and resolution sensitivity for scalable linked visualizations}. In \bibinfo{booktitle}{\emph{Proceedings of the 2019 CHI Conference on Human Factors in Computing Systems}}. \bibinfo{pages}{1--11}.
\newblock


\bibitem[\protect\citeauthoryear{O'Neil, O'Neil, Chen, and Revilak}{O'Neil et~al\mbox{.}}{2009}]%
        {o2007star}
\bibfield{author}{\bibinfo{person}{Patrick O'Neil}, \bibinfo{person}{Elizabeth O'Neil}, \bibinfo{person}{Xuedong Chen}, {and} \bibinfo{person}{Stephen Revilak}.} \bibinfo{year}{2009}\natexlab{}.
\newblock \showarticletitle{The Star Schema Benchmark and Augmented Fact Table Indexing}. In \bibinfo{booktitle}{\emph{Performance Evaluation and Benchmarking}}, \bibfield{editor}{\bibinfo{person}{Raghunath Nambiar} {and} \bibinfo{person}{Meikel Poess}} (Eds.). \bibinfo{publisher}{Springer Berlin Heidelberg}, \bibinfo{address}{Berlin, Heidelberg}, \bibinfo{pages}{237--252}.
\newblock
\showISBNx{978-3-642-10424-4}


\bibitem[\protect\citeauthoryear{Plaisant, Fekete, and Grinstein}{Plaisant et~al\mbox{.}}{2008}]%
        {plaisant2007promoting}
\bibfield{author}{\bibinfo{person}{Catherine Plaisant}, \bibinfo{person}{Jean-Daniel Fekete}, {and} \bibinfo{person}{Georges Grinstein}.} \bibinfo{year}{2008}\natexlab{}.
\newblock \showarticletitle{Promoting Insight-Based Evaluation of Visualizations: From Contest to Benchmark Repository}.
\newblock \bibinfo{journal}{\emph{IEEE TVCG}} \bibinfo{volume}{14}, \bibinfo{number}{1} (\bibinfo{year}{2008}), \bibinfo{pages}{120--134}.
\newblock
\urldef\tempurl%
\url{https://doi.org/10.1109/TVCG.2007.70412}
\showDOI{\tempurl}


\bibitem[\protect\citeauthoryear{Rahman, Aliakbarpour, Kong, Blais, Karahalios, Parameswaran, and Rubinfield}{Rahman et~al\mbox{.}}{2017}]%
        {rahman2017seen}
\bibfield{author}{\bibinfo{person}{Sajjadur Rahman}, \bibinfo{person}{Maryam Aliakbarpour}, \bibinfo{person}{Ha~Kyung Kong}, \bibinfo{person}{Eric Blais}, \bibinfo{person}{Karrie Karahalios}, \bibinfo{person}{Aditya Parameswaran}, {and} \bibinfo{person}{Ronitt Rubinfield}.} \bibinfo{year}{2017}\natexlab{}.
\newblock \showarticletitle{I've Seen "Enough": Incrementally Improving Visualizations to Support Rapid Decision Making}.
\newblock \bibinfo{journal}{\emph{Proc. VLDB Endow.}} \bibinfo{volume}{10}, \bibinfo{number}{11} (\bibinfo{date}{aug} \bibinfo{year}{2017}), \bibinfo{pages}{1262–1273}.
\newblock
\showISSN{2150-8097}
\urldef\tempurl%
\url{https://doi.org/10.14778/3137628.3137637}
\showDOI{\tempurl}


\bibitem[\protect\citeauthoryear{Rahman, Mack, Bendre, Zhang, Karahalios, and Parameswaran}{Rahman et~al\mbox{.}}{2020}]%
        {rahman2020benchmarking}
\bibfield{author}{\bibinfo{person}{Sajjadur Rahman}, \bibinfo{person}{Kelly Mack}, \bibinfo{person}{Mangesh Bendre}, \bibinfo{person}{Ruilin Zhang}, \bibinfo{person}{Karrie Karahalios}, {and} \bibinfo{person}{Aditya Parameswaran}.} \bibinfo{year}{2020}\natexlab{}.
\newblock \showarticletitle{Benchmarking Spreadsheet Systems}. In \bibinfo{booktitle}{\emph{Proceedings of the 2020 ACM SIGMOD International Conference on Management of Data}} (Portland, OR, USA) \emph{(\bibinfo{series}{SIGMOD '20})}. \bibinfo{publisher}{Association for Computing Machinery}, \bibinfo{address}{New York, NY, USA}, \bibinfo{pages}{1589--1599}.
\newblock
\showISBNx{9781450367356}
\urldef\tempurl%
\url{https://doi.org/10.1145/3318464.3389782}
\showDOI{\tempurl}


\bibitem[\protect\citeauthoryear{Salesforce}{Salesforce}{2022}]%
        {salesforce2022visualize}
\bibfield{author}{\bibinfo{person}{Salesforce}.} \bibinfo{year}{2022}\natexlab{}.
\newblock \bibinfo{title}{{Visualize Your Data With Dashboards and Charts}}.
\newblock \bibinfo{howpublished}{{\url{https://trailhead.salesforce.com/content/learn/modules/reports_dashboards/reports_dashboards_visualizing_data}}}.
\newblock


\bibitem[\protect\citeauthoryear{{Sarikaya}, {Correll}, {Bartram}, {Tory}, and {Fisher}}{{Sarikaya} et~al\mbox{.}}{2019}]%
        {sarikaya2019}
\bibfield{author}{\bibinfo{person}{A. {Sarikaya}}, \bibinfo{person}{M. {Correll}}, \bibinfo{person}{L. {Bartram}}, \bibinfo{person}{M. {Tory}}, {and} \bibinfo{person}{D. {Fisher}}.} \bibinfo{year}{2019}\natexlab{}.
\newblock \showarticletitle{What Do We Talk About When We Talk About Dashboards?}
\newblock \bibinfo{journal}{\emph{IEEE TVCG}} \bibinfo{volume}{25}, \bibinfo{number}{1} (\bibinfo{year}{2019}), \bibinfo{pages}{682--692}.
\newblock
\urldef\tempurl%
\url{https://doi.org/10.1109/TVCG.2018.2864903}
\showDOI{\tempurl}


\bibitem[\protect\citeauthoryear{Satyanarayan, Moritz, Wongsuphasawat, and Heer}{Satyanarayan et~al\mbox{.}}{2017}]%
        {2017-vega-lite}
\bibfield{author}{\bibinfo{person}{Arvind Satyanarayan}, \bibinfo{person}{Dominik Moritz}, \bibinfo{person}{Kanit Wongsuphasawat}, {and} \bibinfo{person}{Jeffrey Heer}.} \bibinfo{year}{2017}\natexlab{}.
\newblock \showarticletitle{Vega-Lite: A Grammar of Interactive Graphics}.
\newblock \bibinfo{journal}{\emph{IEEE Trans. Visualization \& Comp. Graphics (Proc. InfoVis)}} (\bibinfo{year}{2017}).
\newblock
\urldef\tempurl%
\url{http://idl.cs.washington.edu/papers/vega-lite}
\showURL{%
\tempurl}


\bibitem[\protect\citeauthoryear{Shneiderman}{Shneiderman}{[n.d.]}]%
        {shneiderman96eyes}
\bibfield{author}{\bibinfo{person}{Ben Shneiderman}.} \bibinfo{year}{[n.d.]}\natexlab{}.
\newblock \showarticletitle{The Eyes Have It: A Task by Data Type Taxonomy for}. In \bibinfo{booktitle}{\emph{Proceedings of IEEE Symposium on Visual Languages}}, Vol.~\bibinfo{volume}{96}.
\newblock


\bibitem[\protect\citeauthoryear{Siddiqui, Kim, Lee, Karahalios, and Parameswaran}{Siddiqui et~al\mbox{.}}{2016}]%
        {siddiqui2016effortless}
\bibfield{author}{\bibinfo{person}{Tarique Siddiqui}, \bibinfo{person}{Albert Kim}, \bibinfo{person}{John Lee}, \bibinfo{person}{Karrie Karahalios}, {and} \bibinfo{person}{Aditya Parameswaran}.} \bibinfo{year}{2016}\natexlab{}.
\newblock \showarticletitle{Effortless Data Exploration with zenvisage: An Expressive and Interactive Visual Analytics System}.
\newblock \bibinfo{journal}{\emph{Proceedings of the VLDB Endowment}} \bibinfo{volume}{10}, \bibinfo{number}{4} (\bibinfo{year}{2016}).
\newblock


\bibitem[\protect\citeauthoryear{Stolte, Tang, and Hanrahan}{Stolte et~al\mbox{.}}{2002}]%
        {stolte2002polaris}
\bibfield{author}{\bibinfo{person}{Chris Stolte}, \bibinfo{person}{Diane Tang}, {and} \bibinfo{person}{Pat Hanrahan}.} \bibinfo{year}{2002}\natexlab{}.
\newblock \showarticletitle{Polaris: A system for query, analysis, and visualization of multidimensional relational databases}.
\newblock \bibinfo{journal}{\emph{IEEE TVCG}} \bibinfo{volume}{8}, \bibinfo{number}{1} (\bibinfo{year}{2002}), \bibinfo{pages}{52--65}.
\newblock


\bibitem[\protect\citeauthoryear{{Tableau Software}}{{Tableau Software}}{2022}]%
        {tableau2022tableau}
\bibfield{author}{\bibinfo{person}{{Tableau Software}}.} \bibinfo{year}{2022}\natexlab{}.
\newblock \bibinfo{title}{{Tableau Desktop | Connect, analyze, and visualize any data}}.
\newblock \bibinfo{howpublished}{{\url{https://www.tableau.com/products/desktop}}}.
\newblock


\bibitem[\protect\citeauthoryear{{TPC-DS}}{{TPC-DS}}{2022}]%
        {tpcds}
{TPC-DS} \bibinfo{year}{2022}\natexlab{}.
\newblock \bibinfo{title}{{TPC-DS}}.
\newblock \bibinfo{howpublished}{{\url{https://www.tpc.org/tpcds/}}}.
\newblock
\newblock
\shownote{Accessed: 2022-10-05.}


\bibitem[\protect\citeauthoryear{{TPC-H}}{{TPC-H}}{2022}]%
        {tpch}
{TPC-H} \bibinfo{year}{2022}\natexlab{}.
\newblock \bibinfo{title}{{TPC-H}}.
\newblock \bibinfo{howpublished}{{\url{http://www.tpc.org/tpch/}}}.
\newblock
\newblock
\shownote{Accessed: 2022-10-05.}


\bibitem[\protect\citeauthoryear{Vogelsgesang, Haubenschild, Finis, Kemper, Leis, Muehlbauer, Neumann, and Then}{Vogelsgesang et~al\mbox{.}}{2018}]%
        {vogelsgesang2018get}
\bibfield{author}{\bibinfo{person}{Adrian Vogelsgesang}, \bibinfo{person}{Michael Haubenschild}, \bibinfo{person}{Jan Finis}, \bibinfo{person}{Alfons Kemper}, \bibinfo{person}{Viktor Leis}, \bibinfo{person}{Tobias Muehlbauer}, \bibinfo{person}{Thomas Neumann}, {and} \bibinfo{person}{Manuel Then}.} \bibinfo{year}{2018}\natexlab{}.
\newblock \showarticletitle{Get Real: How Benchmarks Fail to Represent the Real World}. In \bibinfo{booktitle}{\emph{Proceedings of the Workshop on Testing Database Systems}} (Houston, TX, USA) \emph{(\bibinfo{series}{DBTest'18})}. \bibinfo{publisher}{Association for Computing Machinery}, \bibinfo{address}{New York, NY, USA}, Article \bibinfo{articleno}{1}, \bibinfo{numpages}{6}~pages.
\newblock
\showISBNx{9781450358262}
\urldef\tempurl%
\url{https://doi.org/10.1145/3209950.3209952}
\showDOI{\tempurl}


\bibitem[\protect\citeauthoryear{Wongsuphasawat, Qu, Moritz, Chang, Ouk, Anand, Mackinlay, Howe, and Heer}{Wongsuphasawat et~al\mbox{.}}{2017}]%
        {wongsuphasawat2017voyager}
\bibfield{author}{\bibinfo{person}{Kanit Wongsuphasawat}, \bibinfo{person}{Zening Qu}, \bibinfo{person}{Dominik Moritz}, \bibinfo{person}{Riley Chang}, \bibinfo{person}{Felix Ouk}, \bibinfo{person}{Anushka Anand}, \bibinfo{person}{Jock Mackinlay}, \bibinfo{person}{Bill Howe}, {and} \bibinfo{person}{Jeffrey Heer}.} \bibinfo{year}{2017}\natexlab{}.
\newblock \showarticletitle{Voyager 2: Augmenting visual analysis with partial view specifications}. In \bibinfo{booktitle}{\emph{Proceedings of the 2017 CHI Conference on Human Factors in Computing Systems}}. \bibinfo{pages}{2648--2659}.
\newblock


\bibitem[\protect\citeauthoryear{Yan, Gu, and Rzeszotarski}{Yan et~al\mbox{.}}{2021}]%
        {yan2021tessera}
\bibfield{author}{\bibinfo{person}{Jing~Nathan Yan}, \bibinfo{person}{Ziwei Gu}, {and} \bibinfo{person}{Jeffrey~M Rzeszotarski}.} \bibinfo{year}{2021}\natexlab{}.
\newblock \showarticletitle{Tessera: Discretizing Data Analysis Workflows on a Task Level}. In \bibinfo{booktitle}{\emph{Proceedings of the 2021 CHI Conference on Human Factors in Computing Systems}} (Yokohama, Japan) \emph{(\bibinfo{series}{CHI '21})}. \bibinfo{publisher}{Association for Computing Machinery}, \bibinfo{address}{New York, NY, USA}, Article \bibinfo{articleno}{20}, \bibinfo{numpages}{15}~pages.
\newblock
\showISBNx{9781450380966}
\urldef\tempurl%
\url{https://doi.org/10.1145/3411764.3445728}
\showDOI{\tempurl}


\bibitem[\protect\citeauthoryear{Zeng, Moh, Du, Hoffswell, Lee, Malik, Koh, and Battle}{Zeng et~al\mbox{.}}{2022}]%
        {zeng2022evaluation}
\bibfield{author}{\bibinfo{person}{Zehua Zeng}, \bibinfo{person}{Phoebe Moh}, \bibinfo{person}{Fan Du}, \bibinfo{person}{Jane Hoffswell}, \bibinfo{person}{Tak~Yeon Lee}, \bibinfo{person}{Sana Malik}, \bibinfo{person}{Eunyee Koh}, {and} \bibinfo{person}{Leilani Battle}.} \bibinfo{year}{2022}\natexlab{}.
\newblock \showarticletitle{An Evaluation-Focused Framework for Visualization Recommendation Algorithms}.
\newblock \bibinfo{journal}{\emph{IEEE Transactions on Visualization and Computer Graphics}} \bibinfo{volume}{28}, \bibinfo{number}{1} (\bibinfo{year}{2022}), \bibinfo{pages}{346--356}.
\newblock
\urldef\tempurl%
\url{https://doi.org/10.1109/TVCG.2021.3114814}
\showDOI{\tempurl}


\bibitem[\protect\citeauthoryear{Zgraggen, Galakatos, Crotty, Fekete, and Kraska}{Zgraggen et~al\mbox{.}}{2016}]%
        {zgraggen2016progressive}
\bibfield{author}{\bibinfo{person}{Emanuel Zgraggen}, \bibinfo{person}{Alex Galakatos}, \bibinfo{person}{Andrew Crotty}, \bibinfo{person}{Jean-Daniel Fekete}, {and} \bibinfo{person}{Tim Kraska}.} \bibinfo{year}{2016}\natexlab{}.
\newblock \showarticletitle{How Progressive Visualizations Affect Exploratory Analysis}.
\newblock \bibinfo{journal}{\emph{IEEE TVCG}} (\bibinfo{year}{2016}).
\newblock


\bibitem[\protect\citeauthoryear{Zgraggen, Zhao, Zeleznik, and Kraska}{Zgraggen et~al\mbox{.}}{2018}]%
        {zgraggen2018investigating}
\bibfield{author}{\bibinfo{person}{Emanuel Zgraggen}, \bibinfo{person}{Zheguang Zhao}, \bibinfo{person}{Robert Zeleznik}, {and} \bibinfo{person}{Tim Kraska}.} \bibinfo{year}{2018}\natexlab{}.
\newblock \showarticletitle{Investigating the effect of the multiple comparisons problem in visual analysis}. In \bibinfo{booktitle}{\emph{Proceedings of the 2018 chi conference on human factors in computing systems}}. \bibinfo{pages}{1--12}.
\newblock


\bibitem[\protect\citeauthoryear{Zhang, Zhang, Sellam, and Wu}{Zhang et~al\mbox{.}}{2019}]%
        {zhang2019mining}
\bibfield{author}{\bibinfo{person}{Qianrui Zhang}, \bibinfo{person}{Haoci Zhang}, \bibinfo{person}{Thibault Sellam}, {and} \bibinfo{person}{Eugene Wu}.} \bibinfo{year}{2019}\natexlab{}.
\newblock \showarticletitle{Mining Precision Interfaces From Query Logs}. In \bibinfo{booktitle}{\emph{Proceedings of the 2019 International Conference on Management of Data}} (Amsterdam, Netherlands) \emph{(\bibinfo{series}{SIGMOD '19})}. \bibinfo{publisher}{Association for Computing Machinery}, \bibinfo{address}{New York, NY, USA}, \bibinfo{pages}{988–1005}.
\newblock
\showISBNx{9781450356435}
\urldef\tempurl%
\url{https://doi.org/10.1145/3299869.3319872}
\showDOI{\tempurl}


\bibitem[\protect\citeauthoryear{{Zhicheng Liu} and {Jeffrey Heer}}{{Zhicheng Liu} and {Jeffrey Heer}}{2014}]%
        {Liu14}
\bibfield{author}{\bibinfo{person}{{Zhicheng Liu}} {and} \bibinfo{person}{{Jeffrey Heer}}.} \bibinfo{year}{2014}\natexlab{}.
\newblock \showarticletitle{The Effects of Interactive Latency on Exploratory Visual Analysis}.
\newblock \bibinfo{journal}{\emph{IEEE TVCG}} \bibinfo{volume}{20}, \bibinfo{number}{12} (\bibinfo{year}{2014}), \bibinfo{pages}{2122--2131}.
\newblock


\bibitem[\protect\citeauthoryear{Zhou, Arulraj, Navathe, Harris, and Wu}{Zhou et~al\mbox{.}}{2020}]%
        {zhou2020spes}
\bibfield{author}{\bibinfo{person}{Qi Zhou}, \bibinfo{person}{Joy Arulraj}, \bibinfo{person}{Shamkant Navathe}, \bibinfo{person}{William Harris}, {and} \bibinfo{person}{Jinpeng Wu}.} \bibinfo{year}{2020}\natexlab{}.
\newblock \showarticletitle{SPES: A Two-Stage Query Equivalence Verifier}.
\newblock  (\bibinfo{year}{2020}).
\newblock


\end{thebibliography}

\end{document}